\documentclass[prb, twocolumn, superscriptaddress, footinbib, floatfix, amsmath, amsfonts, amssymb, amsthm, nobalancelastpage]{revtex4-2}

\usepackage[utf8]{inputenc}
\usepackage[T1]{fontenc}
\usepackage[main=english]{babel}

\usepackage[table]{xcolor}
\usepackage{graphicx}
\usepackage[caption=false]{subfig}

\usepackage{siunitx}
\DeclareSIUnit\angstrom{\text{Å}}
\usepackage[version=4]{mhchem}

\usepackage{bm}
\usepackage{bbm}
\usepackage{upgreek}

\usepackage{hyperref}
\hypersetup{colorlinks=true, linkcolor=blue, citecolor=blue, urlcolor=blue, filecolor=red, pdfstartview=}

\newcommand*{\R}{\mathbb{R}}
\newcommand*{\iu}{\mathrm{i}}
\newcommand*{\Elr}{\mathrm{e}}
\newcommand*{\one}{\mathbbm{1}}
\newcommand*{\Pauli}[1]{\upsigma_{#1}}

\newcommand*{\abs}[1]{\left\lvert {#1} \right\rvert}
\newcommand*{\dd}[2][]{\mathop{}\!\mathrm{d}^{#1} {#2}}
\newcommand*{\var}[2][]{\mathop{}\!\delta_{#1} {#2}}

\newcommand*{\vdot}{\bm{\cdot}}
\newcommand*{\grad}{\bm{\nabla}}
\newcommand*{\vb}[1]{\bm{#1}}
\newcommand*{\vu}[1]{\bm{\hat{#1}}}

\newcommand*{\Kd}{\updelta}
\DeclareMathOperator{\Dd}{\updelta}

\DeclareMathOperator{\tr}{tr}
\DeclareMathOperator{\diag}{diag}
\DeclareMathOperator{\arcsinh}{arcsinh}

\newcommand*{\hc}{\mathrm{h.c.}}

\makeatletter
\newcommand{\thickhline}{%
    \noalign {\ifnum 0=`}\fi \hrule height 1pt
    \futurelet \reserved@a \@xhline
}
\newcolumntype{"}{@{\hskip\tabcolsep\vrule width 1pt\hskip\tabcolsep}}
\makeatother

\begin{document}
\title{Constraints on the superconducting state of \ce{Sr2RuO4} from elastocaloric measurements}
\author{Grgur Palle}
\email{grgur.palle@kit.edu}
\affiliation{Institute for Theory of Condensed Matter, Karlsruhe Institute of Technology, Karlsruhe, Germany}
\author{Clifford Hicks}
\affiliation{School of Physics and Astronomy, University of Birmingham, Birmingham, United Kingdom}
\affiliation{Max Planck Institute for Chemical Physics of Solids, Dresden, Germany}
\author{Roser Valentí}
\affiliation{Institute for Theoretical Physics, Goethe-University Frankfurt, Frankfurt am Main, Germany}
\author{Zhenhai Hu}
\affiliation{Max Planck Institute for Chemical Physics of Solids, Dresden, Germany}
\author{You-Sheng Li}
\affiliation{Max Planck Institute for Chemical Physics of Solids, Dresden, Germany}
\author{Andreas Rost}
\affiliation{Scottish Universities Physics Alliance, School of Physics and Astronomy, University of St Andrews, St Andrews, United Kingdom}
\author{Michael Nicklas}
\affiliation{Max Planck Institute for Chemical Physics of Solids, Dresden, Germany}
\author{Andrew P. Mackenzie}
\affiliation{Max Planck Institute for Chemical Physics of Solids, Dresden, Germany}
\affiliation{Scottish Universities Physics Alliance, School of Physics and Astronomy, University of St Andrews, St Andrews, United Kingdom}
\author{Jörg Schmalian}
\affiliation{Institute for Theory of Condensed Matter, Karlsruhe Institute of Technology, Karlsruhe, Germany}
\affiliation{Institute for Quantum Materials and Technologies, Karlsruhe Institute of Technology, Karlsruhe, Germany}
\date{\today}
\begin{abstract}
Strontium ruthenate \ce{Sr2RuO4} is an unconventional superconductor whose pairing symmetry has not been fully clarified, despite more than two decades of intensive research.
Recent NMR Knight shift experiments have rekindled the \ce{Sr2RuO4} pairing debate by giving strong evidence against all odd-parity pairing states, including chiral $p$-wave pairing that was for a long time the leading pairing candidate.
Here, we exclude additional pairing states by analyzing recent elastocaloric measurements [YS.~Li \textit{et al.}, \href{https://doi.org/10.1038/s41586-022-04820-z}{Nature \textbf{607}, 276--280 (2022)}].
To be able to explain the elastocaloric experiment, we find that unconventional even-parity pairings must include either large $d_{x^2 - y^2}$-wave or large $\{d_{xz} \mid d_{yz}\}$-wave admixtures, where the latter possibility arises because of the body-centered point group symmetry.
These $\{d_{xz} \mid d_{yz}\}$-wave admixtures take the form of distinctively body-centered-periodic harmonics that have horizontal line nodes.
Hence $g_{xy(x^2-y^2)}$-wave and $d_{xy}$-wave pairings are excluded as possible dominant even pairing states.
\end{abstract}

\maketitle

\section{Introduction}
The nature of the superconductivity of strontium ruthenate (SRO) remains elusive. In the three decades following its discovery~\cite{Maeno1994}, an impressive array of experiments have been performed with high precision and on exceedingly pure samples~\cite{Mackenzie2003, Maeno2011, Kallin2012, LiuMao2015, Mackenzie2017}.
Yet the most straightforward interpretations of the various experimental results are regularly at odds with one another.
Although many proposals~\cite{Kivelson2020, Romer2020, Willa2021, Yuan2021, Sheng2022, Yuan2023,
Wagner2021,
Clepkens2021, Romer2021,
Suh2020, Fukaya2022,
Leggett2021,
Huang2021-p2,
Gingras2022, Scaffidi2023} have been made on how the assortment of experimental results might be reconciled, no consensus has formed around which proposal is the correct one.
Before presenting our results, in the next six paragraphs we review what is currently known about the pairing state. This literature review is not essential to our argument and can be skipped.

The superconductivity (SC) of SRO is unconventional.
This has been established early on by the absence of a Hebel-Slichter peak~\cite{Hebel1957, *Hebel1959} in the NMR relaxation rate $1/T_1$~\cite{Ishida1997, Ishida2000, Murakawa2007}, and
by the large suppression of the SC transition temperature $T_c$ by non-magnetic impurities~\cite{Mackenzie1998, *Mackenzie1998-E, Mao1999, Kikugawa2002, Kikugawa2004} that saturates the Abrikosov-Gor'kov bound~\cite{Abrikosov1961, Gorkov2008}.
Subsequent experiments have only further confirmed the unconventional character of SRO's SC.

The pairing of SRO is more likely to be even than not.
Recent\textsuperscript{\citenum{Note1}} NMR Knight shift~\cite{Pustogow2019, Ishida2020, Chronister2021} and polarized neutron scattering~\cite{Petsch2020} experiments strongly favor singlet pairing,
as do numerous studies~\cite{Mackenzie2017, Note2} indicating that the in-plane critical field $B_{c2 \parallel ab}$ is Pauli limited~\cite{Clogston1962}.
Although the observation of $\pi$ phase shifts~\cite{Nelson2004} and half-quantum vortices~\cite{Jang2011, Yasui2017, Cai2022} is at tension with even-parity SC, possible explanations do exist~\cite{Yuan2021, Zutic2005, Lindquist2023}.
Reconciling an \SI{80}{\percent} drop in the in-plane Knight shift~\cite{Chronister2021} with triplet pairing, or a strained critical field anisotropy $B_{c2 \parallel ab} / B_{c2 \parallel c} \sim 3$~\cite{Steppke2017} far below the SC anisotropy $\xi_{ab} / \xi_c \sim 60$~\cite{Rastovski2013, Kittaka2014} without Pauli limiting~\cite{Mackenzie2017}, is significantly more challenging, but perhaps possible~\cite{Ramires2016, Ramires2017}.

\footnotetext[1]{ 
The heating caused by NMR pulses~\cite{Pustogow2019, Ishida2020} has rendered early NMR Knight shift experiments~\cite{Ishida1998}, nicely summarized in Figure~14 of Ref.~\cite{Murakawa2007}, invalid.
The NMR pulse heat-up effect acts on a time scale much shorter than $T_1$ and has not invalidated the early NMR relaxation rate studies~\cite{Pustogow2019}.
An early polarized neutron scattering study~\cite{Duffy2000} has been superseded by a new one~\cite{Petsch2020} with better statistics, carried out at a smaller magnetic field.
}

\footnotetext[2]{ 
The evidence for a Pauli-limited $B_{c2 \parallel ab}$ is threefold:
(i) the SC-normal state transition is first-order below $0.5 T_c$, as seen in the hysteresis~\cite{Yonezawa2013, Yonezawa2014, Kittaka2014} and jumps in the specific heat~\cite{Deguchi2002, Yonezawa2014}, thermal conductivity~\cite{Deguchi2002}, magnetocaloric effect~\cite{Yonezawa2013}, ac magnetic susceptibility~\cite{Yaguchi2002}, magnetization~\cite{Kittaka2014}, and Knight shift~\cite{Chronister2021};
(ii) the measured intrinsic SC anisotropy $\xi_{ab} / \xi_c \sim 60$~\cite{Rastovski2013, Kittaka2014} exceeds the critical field anisotropy $B_{c2 \parallel ab} / B_{c2 \parallel c} \sim 20$~\cite{Kittaka2009} by a factor of $3$ in the unstrained case, and by a factor of $20$ under $\langle 100 \rangle$ uniaxial pressure that maximally enhances $T_c$~\cite{Steppke2017},
whereas for orbitally limited $B_{c2 \parallel ab}$ the two ratios would be comparable; and
(iii) $B_{c2 \parallel ab} \propto \Delta / \mu_B \propto T_c$ under small uniaxial strain~\cite{Jerzembeck2023}, as expected for Pauli limiting.
}

The evidence for time-reversal symmetry breaking (TRSB) is mixed.
Zero-field muon spin relaxation (ZF-$\mu$SR)~\cite{Luke1998, Luke2000, Higemoto2014, Grinenko2021-unaxial, Grinenko2021-isotropic} and polar Kerr effect~\cite{Xia2006, Kapitulnik2009} experiments indicate TRSB at a $T_{\text{TRSB}}$ at or very near $T_c$,
yet the current response of micron-sized Josephson junctions~\cite{[{}][{ Note: contrary to what they say, the inversion symmetry $I_c^{+}(H) = - I_c^{-}(-H)$ that becomes restored for small junctions is precisely time-reversal symmetry.}]{Saitoh2015}, Kashiwaya2019}
exhibits time-reversal invariance.
Under $\langle 100 \rangle$ uniaxial pressure, ZF-$\mu$SR~\cite{Grinenko2021-unaxial} observes a large splitting between $T_{\text{TRSB}}$ and $T_c$,\textsuperscript{\citenum{Note3}} yet no signatures of a TRSB phase transition below $T_c$ have been found in heat capacity~\cite{Li2021} or elastocaloric~\cite{Li2022} measurements.
Under disorder and hydrostatic pressure, no splitting between SC and TRSB is observed in ZF-$\mu$SR~\cite{Grinenko2021-isotropic}.
Preliminary ZF-$\mu$SR measurements point towards splitting of SC and TRSB under $\langle 110 \rangle$ uniaxial stress~\cite{Grinenko2023}.
In the presence of TRSB, spontaneous magnetization and currents are generically expected to appear around domain walls, edges, and defects, yet scanning SQUID and Hall probe microscopy~\cite{Tamegai2003, Bjornsson2005, Kirtley2007, Hicks2010, Curran2011, Curran2014, Mueller2023, Curran2023} has failed to find any evidence for them.
Josephson junction experiments~\cite{Saitoh2015, Kidwingira2006, Anwar2013, Anwar2017} show signs of SC domains in their interference patters, switching behavior, and size dependence of their transport properties, but the domains themselves need not be chiral.

\footnotetext[3]{ 
In one sample~\cite{Grinenko2021-unaxial}, $T_{\text{TRSB}}$ and $T_c$ split even without any external pressure.
}

The coupling of SC to strain is partially known from measurements of elastic constants.
The main obstacle to making these measurements conclusive is the fact that strain inhomogeneities, such as stacking faults or lattice dislocations, mix elastic waves of different symmetry.\textsuperscript{\citenum{Note4}}
That said, according to elastic constant measurements, the SC order appears to couple quadratically to $\varepsilon_{xx} - \varepsilon_{yy} \in B_{1g}$ strain and possibly linearly to $\varepsilon_{xy} \in B_{2g}$ strain.
(Irreducible representations (irreps) of SRO are summarized in Table~\ref{tab:ex-D4h-func}.)
The evidence for the former is the quadratic dependence of $T_c$ on $\varepsilon_{xx} - \varepsilon_{yy}$, whether measured globally~\cite{Hicks2014, Steppke2017, Barber2019} or locally~\cite{Watson2018}, and the absence of a jump at $T_c$ in the shear elastic modulus $C_{B_{1g}} = \tfrac{1}{2} (C_{11} - C_{12})$~\cite{Matsui2001, Benhabib2021, Ghosh2021}.
The evidence for the latter is a jump at $T_c$ in the shear elastic constant $C_{66} \in B_{2g}$~\cite{Okuda2003, Benhabib2021, Ghosh2021}, as measured by ultrasound.
However, the magnitude of this jump varies by a factor of $50$ between the two experimental groups~\cite{Benhabib2021, Ghosh2021} and direct measurements of $T_c$ under $[110]$ strain show linear dependence without any splitting and whose magnitude can be fully accounted without any linear coupling to $\varepsilon_{xy}$~\cite{Jerzembeck2023-unpublished}.
This raises the possibility that the observed jump in $C_{66}$ is due to lattice defect effects that, however, need to be channel selective so as to not generate a jump in $C_{B_{1g}}$. One such proposal~\cite{Willa2021} is that a subleading pairing channel activates near dislocations; the product of the leading and subleading pairing irreps then determines which elastic modulus experiences a jump.
No jump has been observed for the elastic modulus $C_{44} \in E_g$~\cite{Matsui2001, Ghosh2021}, indicating that the coupling to $E_g$ strain is quadratic.
Large jumps in the $A_{1g}$ components of the viscosity tensor have recently been discovered at $T_c$~\cite{Ghosh2022}.

\footnotetext[4]{ 
As pointed out in~\cite{Willa2021}, dislocations give contributions to elastic constants that are on the order of \SI{1}{\percent}, which is two orders of magnitude larger than the (larger of the two sets of) measured jumps of the elastic constants at $T_c$~\cite{Ghosh2021}.
}

\begin{table}[t]
\caption{Examples of functions transforming according to the irreps of the point group $D_{4h}$ of SRO.
$D_{4h}$ is generated by fourfold rotations around $z$, twofold rotations around $x$ and $y$, twofold rotations around the diagonals $x \pm y$, and parity.
It has five even ($A_{1g}$, $A_{2g}$, $B_{1g}$, $B_{2g}$, $E_g$) and five odd ($A_{1u}$, $A_{2u}$, $B_{1u}$, $B_{2u}$, $E_u$) irreps, of which $E_g$ and $E_u$ are 2D.}
{\renewcommand{\arraystretch}{1.3}
\renewcommand{\tabcolsep}{2.8pt}
\begin{tabular}{c|c|c|c|c} \hline\hline
$A_{1g}$ & $A_{2g}$ & $B_{1g}$ & $B_{2g}$ & $E_g$ \\
$1$, $x^2+y^2$, $z^2$ & $xy(x^2-y^2)$ & $x^2-y^2$ & $xy$ & $\{yz \mid -xz\}$ \\ \hline
$A_{1u}$ & $A_{2u}$ & $B_{1u}$ & $B_{2u}$ & $E_u$ \\
$xyz(x^2-y^2)$ & $z$ & $xyz$ & $(x^2-y^2)z$ & $\{x \mid y\}$
\\ \hline\hline
\end{tabular}}
\label{tab:ex-D4h-func}
\end{table}

The preponderance of evidence points towards line nodes.
The expected dependence on temperature is found in the heat capacity~\cite{NishiZaki2000, Deguchi2004, Kittaka2018}, ultrasound attenuation rate~\cite{Matsui2001, Lupien2001}, NMR relaxation rate~\cite{Ishida2000}, and London penetration depth~\cite{Bonalde2000}.
In weak in-plane fields, the heat capacity~\cite{Deguchi2004-p2, Kittaka2018} and Knight shift~\cite{Chronister2021} obey Volovik scaling ($\propto \sqrt{B / B_{c2}}$) expected of line nodes~\cite{Volovik1993}.
The in-plane thermal conductivity~\cite{Suzuki2002, Hassinger2017} exhibits universal transport, which is a type of transport found only in nodal SC~\cite{Lee1993, Balatsky1995, Sun1995, Graf1996}.
Finally, STM spectroscopy~\cite{Firmo2013, Sharma2020} shows a $V$-shaped conductance minimum.\textsuperscript{\citenum{Note5}}
The only evidence to the contrary is an STM/S study~\cite{Suderow2009} that scanned micron-sized grains ($\sim 10 \, \xi_{ab}$) situated on top of SC aluminium and found an implausibly large SC gap $\Delta$ of \SI{3.5}{\kelvin}.
Given that so many studies~\cite{NishiZaki2000, Deguchi2004, Kittaka2018, Matsui2001, Lupien2001, Ishida2000, Bonalde2000, Deguchi2004-p2} found nodal behavior, in some cases down to as low as $\SI{0.04}{\kelvin} \approx T_c / 30$, any fully gapped SC must have extraordinarily deep minima.

\footnotetext[5]{ 
One should keep in mind that STM mostly probes the $\alpha, \beta$ bands because of their $d_{xz}, d_{yz}$ orbital characters which make their overlaps with the tip (along $z$) large.
}

The location and orientation of the line node(s) is not settled.
Heat capacity~\cite{Kittaka2018} and in-plane thermal conductivity~\cite{Tanatar2001-p2, Izawa2001} both display a fourfold anisotropy in their dependence on the in-plane $\vb{B}$ orientation.\textsuperscript{\citenum{Note6}}
Since these anisotropies are small ($\sim \SI{1}{\percent}$), they can be explained by both horizontal and vertical nodes.
That the heat capacity anisotropy has the same sign down to $T_c / 20$ appears to exclude $d_{xy}$-wave pairing~\cite{Kittaka2018}, and maybe other pairings too.
The universal heat transport along $c$ has been found finite with $2 \sigma$ significance~\cite{Hassinger2017}, indicating that nodal quasi-particles have a finite $c$-axis velocity.
If true, this result is strong evidence against symmetry-enforced horizontal line nodes.
A resonance at transfer energy $\approx 2 \Delta$ and momentum with a finite $z$ component was reported below $T_c$ in the inelastic neutron scattering intensity~\cite{Iida2020}, suggesting horizontal line nodes, but was not reproduced in subsequent measurements~\cite{Jenni2021}.
In the Fourier transform of the real-space STM tunneling conductance~\cite{Sharma2020}, peaks were found at nesting vectors expected of $d_{x^2-y^2}$-wave SC.
However, the peaks are not clearly resolved because of noise and compatibility with other pairings was not investigated.

\footnotetext[6]{ 
As pointed out in~\cite{Kittaka2018}, little useful information can be extracted from the out-of-plane field-angle anisotropy.
}

Compelling evidence on SRO's gap structure has recently emerged from measurements performed under uniaxial pressure.
When $\langle 100 \rangle$ uniaxial pressure is applied on SRO, its SC is drastically enhanced~\cite{Hicks2014, Taniguchi2015, Steppke2017, Barber2019, Jerzembeck2023}, with $T_c$ increasing from \SI{1.5}{\kelvin} to a maximal \SI{3.5}{\kelvin} before decaying again.
The most likely cause of this enhancement is the Lifshitz transition that occurs at $\varepsilon_{xx} = \SI{-0.44}{\percent} \equiv \varepsilon_{\text{VH}}$ strain~\cite{Steppke2017, Barber2019, Sunko2019} and is accompanied by an increase in the density of states (DOS).
The DOS peaks at $\varepsilon_{\text{VH}}$, as does the normal-state entropy~\cite{Li2022}.
In the SC state, however, the entropy becomes a \emph{minimum} at $\varepsilon_{\text{VH}}$, as directly measured by the elastocaloric effect~\cite{Li2022}.
As we later explain, this is only possible if SRO's SC does not have vertical line nodes at the Van Hove lines that induce the DOS peak at $\varepsilon_{\text{VH}}$.
This is a severe constraint on possible pairing states, one whose implications we explore in this article.
The final piece of the argument is that these properties of strained SRO carry over to the unstrained SC state, which is supported by the absence of any signatures of a bulk SC state change at finite strain in the heat capacity~\cite{Li2021}, elastocaloric effect~\cite{Li2022}, or NMR Knight shift~\cite{Pustogow2019, Chronister2021}.

The main result of this work is that, among even pairings, only $s$-wave ($A_{1g}$), $d_{x^2 - y^2}$-wave ($B_{1g}$), and body-centered periodic $\{d_{xz} \mid d_{yz}\}$-wave ($E_g$) pairings gap the Van Hove lines.
Thus the SC state must include admixtures from at least one of these three pairings to be consistent with the elastocaloric experiment.
The logic of our argument does not put any constraints on the subleading channels.
For instance, almost degenerate states like $d_{x^2 - y^2} + \iu \, g_{xy(x^2 - y^2)}$~\cite{Kivelson2020} or $s' + \iu \, d_{xy}$~\cite{Clepkens2021, Romer2021} are consistent with a dominant $d_{x^2 - y^2}$-wave or $s$-wave state, respectively.
Among odd-parity pairings, all irreps can gap the Van Hove lines.
However, $A_{2u}$ and $B_{2u}$ pairings must be made of body-centered periodic wavefunctions, and for the rest we find non-trivial constraints on the orientations of their Balian-Werthamer $\vb{d}$-vectors~\cite{Balian1963}.

The paper is organized as follows.
In Sec.~\ref{sec:SRO-basics} we review some basic properties of SRO.
After that, in Sec.~\ref{sec:elasto}, we explain what has been measured in the elastocaloric experiment~\cite{Li2022} and why these measurements forbid vertical line nodes at the Van Hove lines.
The precise location of the Van Hove lines is the subject of Sec.~\ref{sec:vH-line}.
Because of its multiband nature, SRO supports a richer set of pairing states than single-band SC~\cite{Ramires2019, Kaba2019, *Kaba2019-E, Huang2019}, which is briefly discussed at the beginning of Sec.~\ref{sec:behavior-vH} and at length in Appendix~\ref{sec:SC-construct}.
Section~\ref{sec:behavior-vH} contains the main results of our work: how the momentum and spin-orbit parts of the SC gap behave near the Van Hove lines and which SC states are excluded by the elastocaloric measurements.
Table~\ref{tab:main-result} is our main result.
In the last section, we discuss our results.

\section{Crystal and electronic structure} \label{sec:SRO-basics}
SRO is a layered perovskite with a body-centered tetragonal lattice ($a = \SI{3.86}{\angstrom}$, $c = \SI{12.7}{\angstrom}$), space group $I4/mmm$, and point group $D_{4h}$~\cite{Mackenzie2003, Bergemann2003}.
The character table of $D_{4h}$ is given in Table~\ref{tab:D4h-char}.

\begin{figure}[t]
\includegraphics[width=0.85\columnwidth]{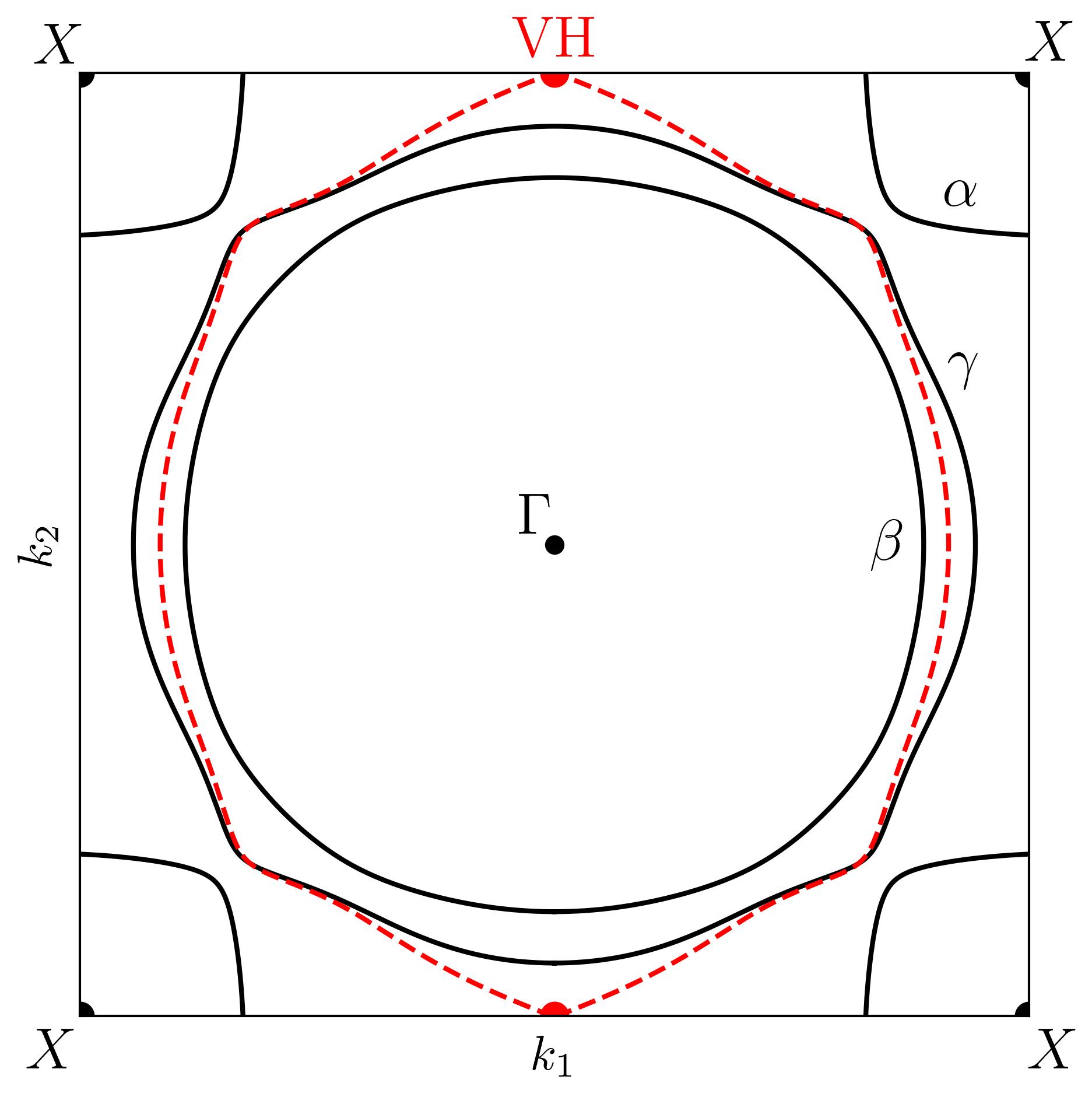}
\caption{The Fermi surfaces of SRO.
The solid black lines are the $k_3 = 0$ cross-sections of the cylindrical $\alpha$, $\beta$, and $\gamma$ Fermi sheets of unstrained SRO, as determined by our tight-binding model [Appendix~\ref{sec:SRO-TBA}].
The dashed red line is the $\gamma$ band of SRO under Van Hove unaxial strain $\varepsilon_{100} = \SI{-0.44}{\percent} \equiv \varepsilon_{\text{VH}}$~\cite{Sunko2019, Li2022}.
At this strain, the $\gamma$ band opens at the Van Hove lines $\left(0, \pm \frac{\pi}{a}, k_3\right)$, here denoted with red dots.}
\label{fig:bands}
\end{figure}

SRO has three conduction bands, conventionally referred to as $\alpha$, $\beta$, and $\gamma$, with cylindrical Fermi sheets~\cite{Mackenzie2003, Bergemann2003}.
They are depicted in Figure~\ref{fig:bands}.
These bands primarily derive from the $t_{2g}$ orbital manifold of the \ce{Ru} atoms, which is made of $d_{yz}$, $d_{zx}$, and $d_{xy}$ orbitals~\cite{Mackenzie2003, Bergemann2003, Gingras2022}.
To a first approximation, due to the high anisotropy, $d_{yz}$ and $d_{zx}$ have 1D tight-binding dispersions:
\begin{align}
\epsilon_{yz}(\vb{k}) &= - \mu - 2 t \cos a k_2, \label{eq:orbit-disp1} \\
\epsilon_{zx}(\vb{k}) &= - \mu - 2 t \cos a k_1, \label{eq:orbit-disp2}
\end{align}
whereas $d_{xy}$ has a 2D tight-binding dispersion:
\begin{align}
\begin{aligned}
\epsilon_{xy}(\vb{k}) &= - \mu - 2 t (\cos a k_1 + \cos a k_2) \\
&\qquad - 4 t' \cos a k_1 \cos a k_2,
\end{aligned} \label{eq:orbit-disp3}
\end{align}
where $(\mu, t, t') \approx (0.35, 0.3, 0.1) \, \si{\electronvolt}$~\cite{Roising2019, Suh2020}.
After introducing interorbital mixing and spin-orbit coupling, $\epsilon_{yz}(\vb{k})$ and $\epsilon_{zx}(\vb{k})$ hybridize into the quasi-1D $\alpha$ and $\beta$ bands, whereas $\epsilon_{xy}(\vb{k})$ hybridizes into the quasi-2D $\gamma$ band [Figure~\ref{fig:bands}].
Interlayer hopping adds warping along $k_3$.
Below \SI{25}{\kelvin}, SRO is a quasi-2D Fermi liquid.
Its quasi-particles are strongly renormalized by electronic correlations~\cite{Mackenzie2003, Tamai2019}.
In the clean limit, SRO develops SC below \SI{1.5}{\kelvin}~\cite{Mackenzie2003}.

\begin{table}[t]
\caption{The character table of the point group $D_{4h}$~\cite{Dresselhaus2007}.
Irreps are divided into even ($g$) and odd ($u$) ones.
$C_4$ are rotations by $\pm \pi/2$ around the $z$ axis.
$C_2$, $C_2'$, and $C_2''$ are rotations by $\pi$ around the $z$ axis, $x$ or $y$ axes, and diagonals $x \pm y$, respectively.
$P$ is parity.
$S_4$, $\Sigma_h$, $\Sigma_v'$, $\Sigma_d''$ are compositions of $C_4$, $C_2$, $C_2'$, $C_2''$ with $P$, respectively.}
{\renewcommand{\arraystretch}{1.3}
\renewcommand{\tabcolsep}{4.2pt}
\begin{tabular}{c|rrrrr|rrrrr} \hline\hline
$D_{4h}$ & $E$ & $2 C_4$ & $C_2$ & $2 C_2'$ & $2 C_2''$ & $P$ & $2 S_4$ & $\Sigma_h$ & $2 \Sigma_v'$ & $2 \Sigma_d''$
\\ \hline
$A_{1g}$ & $1$ & $1$ & $1$ & $1$ & $1$ & $1$ & $1$ & $1$ & $1$ & $1$
\\
$A_{2g}$ & $1$ & $1$ & $1$ & $-1$ & $-1$ & $1$ & $1$ & $1$ & $-1$ & $-1$
\\
$B_{1g}$ & $1$ & $-1$ & $1$ & $1$ & $-1$ & $1$ & $-1$ & $1$ & $1$ & $-1$
\\
$B_{2g}$ & $1$ & $-1$ & $1$ & $-1$ & $1$ & $1$ & $-1$ & $1$ & $-1$ & $1$
\\
$E_g$ & $2$ & $0$ & $-2$ & $0$ & $0$ & $2$ & $0$ & $-2$ & $0$ & $0$
\\ \hline
$A_{1u}$ & $1$ & $1$ & $1$ & $1$ & $1$ & $-1$ & $-1$ & $-1$ & $-1$ & $-1$
\\
$A_{2u}$ & $1$ & $1$ & $1$ & $-1$ & $-1$ & $-1$ & $-1$ & $-1$ & $1$ & $1$
\\
$B_{1u}$ & $1$ & $-1$ & $1$ & $1$ & $-1$ & $-1$ & $1$ & $-1$ & $-1$ & $1$
\\
$B_{2u}$ & $1$ & $-1$ & $1$ & $-1$ & $1$ & $-1$ & $1$ & $-1$ & $1$ & $-1$
\\
$E_u$ & $2$ & $0$ & $-2$ & $0$ & $0$ & $-2$ & $0$ & $2$ & $0$ & $0$
\\ \hline\hline
\end{tabular}}
\label{tab:D4h-char}
\end{table}

Below \SI{25}{\kelvin}, SRO is well-described by a tight-binding model based on the $t_{2g}$ orbitals of ruthenium~\cite{Roising2019, Suh2020, Zabolotnyy2013, Cobo2016}.
Within it, the hopping amplitudes $t_{\vb{\delta}}$ between neighboring lattice sites are significantly constrained by the symmetries of SRO.
In a body-centered lattice, hopping amplitudes along the half-diagonal $\vb{\delta} = \frac{1}{2} \left(a \vu{e}_1 + a \vu{e}_2 + c \vu{e}_3\right)$, as well as many other $\vb{\delta}$, are additionally possible.
However, all such characteristically body-centered hoppings necessarily connect different layers and are thus suppressed by SRO's anisotropy.
For the purpose of making estimates, throughout this paper we employ the normal-state model of Ref.~\cite{Roising2019}, the details of which are provided in Appendix~\ref{sec:SRO-TBA}.

\section{Implications of elastocaloric measurements} \label{sec:elasto}
The elastocaloric effect describes the change in the temperature that accompanies an adiabatic change in the strain $\varepsilon_{\alpha\beta}$.
By measuring it, one may determine the dependence of the entropy $S$ on strain.
This is made possible by the thermodynamic identity:
\begin{equation}
\left.\frac{\partial T}{\partial \varepsilon_{\alpha\beta}}\right|_{S} =  - \frac{T}{C_{\varepsilon}(T)} \left.\frac{\partial S}{\partial \varepsilon_{\alpha\beta}}\right|_{T}, \label{eq:elasto-id}
\end{equation}
where $C_{\varepsilon}(T) = T \left(\partial S / \partial T\right)_{\varepsilon}$ is the heat capacity at constant strain.
Recently, important progress has been made in the experimental techniques for measuring the elastocaloric effect and in their analysis for correlated electron systems~\cite{Ikeda2019, Straquadine2020, Ikeda2021}.

\begin{figure}[t]
\includegraphics[width=\columnwidth]{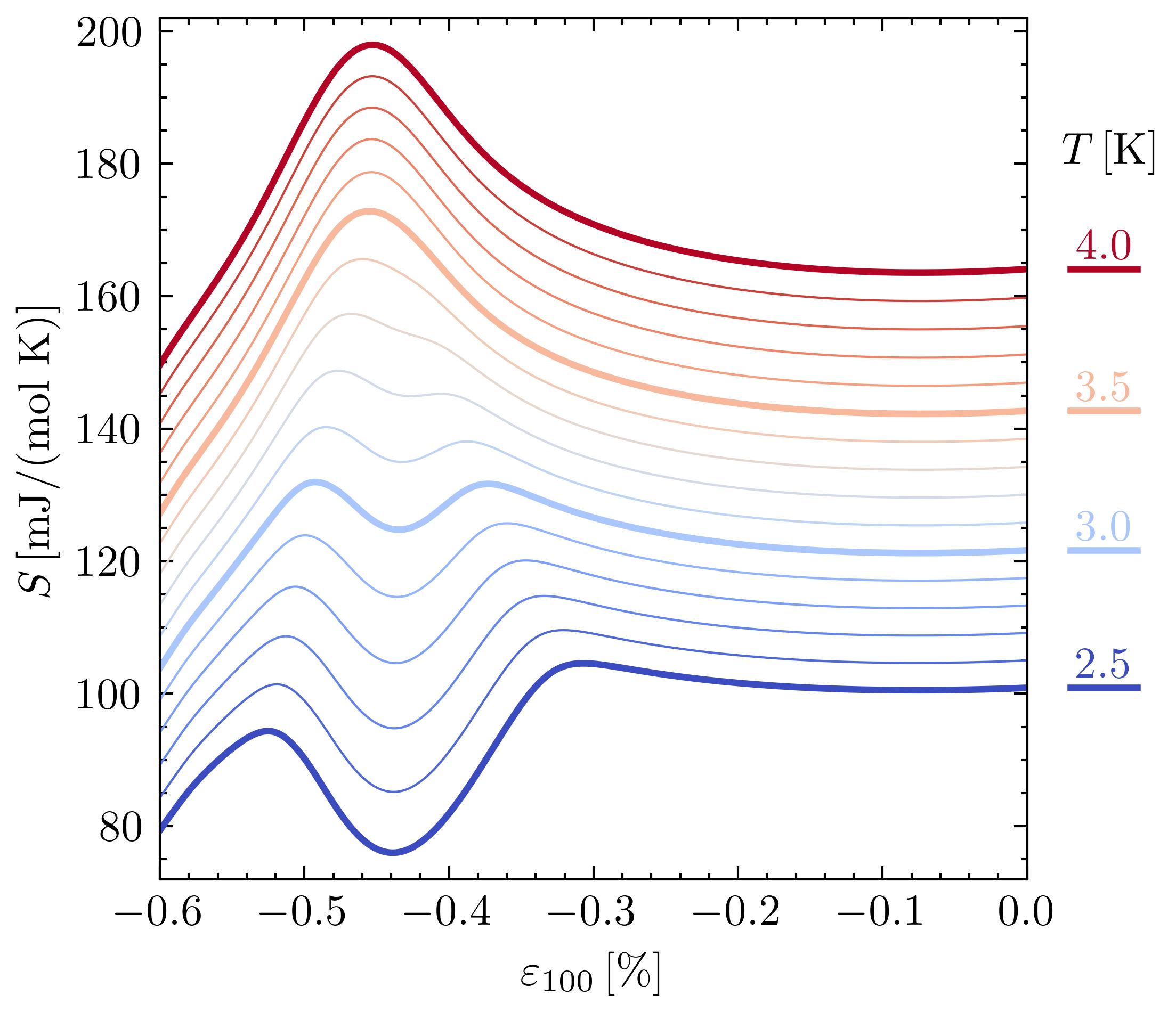}
\caption{The entropy $S$ as a function of strain $\varepsilon_{100}$ at constant temperatures $T$ ranging from \SI{2.5}{\kelvin} (blue) to \SI{4.0}{\kelvin} (red) in \SI{0.1}{\kelvin} increments.
The entropies at different temperatures are naturally offset from each other by their temperature dependence.
The entropy has been reconstructed from elastocaloric measurements~\cite{Li2022} using Eq.~\eqref{eq:elasto-id}.
At Van Hove strain $\varepsilon_{100} = \SI{-0.44}{\percent} \equiv \varepsilon_{\text{VH}}$, $T_c$ attains its maximal value of \SI{3.5}{\kelvin}.
Above (below) \SI{3.5}{\kelvin}, the entropy has a maximum (minimum) at $\varepsilon_{\text{VH}}$ strain.}
\label{fig:elasto}
\end{figure}

The elastocaloric effect has been measured last year for strain applied along the $[100]$ direction~\cite{Li2022}.
Numerical analysis of this dense data set~\cite{Note7} enables the separation of the contribution from $C_{\varepsilon}$ and the reconstruction of the dependence of the entropy on strain; see Figure~\ref{fig:elasto}.

\footnotetext[7]{ 
The elastocaloric data of Ref.~\cite{Li2022} are available at \url{https://doi.org/10.17630/6a4a06c6-38d3-464f-88d1-df8d2dbf1e75}.
}

As clearly seen in the figure, the normal-state entropy has a maximum at the Van Hove strain $\varepsilon_{100} = \SI{-0.44}{\percent} \equiv \varepsilon_{\text{VH}}$.
As we enter the SC state, however, this maximum becomes a \emph{minimum} as a function of strain.
To understand this behavior, let us recall that the entropy of a Fermi liquid is given by~\cite{SolyomVol3, Coleman2015}:
\begin{gather}
S = V \frac{\pi^2}{3} k_B^2 T \int \dd{E} \, g(E) \Dd_{T}(E), \label{eq:Fermi-liquid-entropy} \\
\Dd_{T}(E) = \frac{3}{\pi^2 k_B T} \big[- f \log f - (1-f) \log(1-f)\big],
\end{gather}
where $V$ is the volume, $g(E)$ the DOS, $E$ is relative to the chemical potential, and $f = 1/(\Elr^{E/k_B T} + 1)$.
$\Dd_{T}(E) \to \Dd(E)$ as $T \to 0$ so $S \sim T g(0)$.
This formula applies to both the normal and the SC state.
Thus to understand the entropy, we need to study the DOS near the Fermi level $E = 0$.

In the normal state, at Van Hove strain the $\gamma$ band experiences a Lifshitz transition in which its cylindrical Fermi surface opens at the Van Hove lines $\vb{k}_{\text{VH}} \approx \left(0, \pm \frac{\pi}{a}, k_3\right)$ along the $k_2$-direction~\cite{Steppke2017, Barber2019, Sunko2019}.
This is shown in Figure~\ref{fig:bands}.
Because of the particularly weak $k_3$-dispersion of the $\gamma$ band at $\vb{k}_{\text{VH}}$ ($\sim \SI{1}{\kelvin}$), the Van Hove lines contribute a pronounced peak in the DOS that is only rounded on an energy scale of about one kelvin~\cite{Li2022}.
It is this peak in the DOS that explains the observed normal-state entropy maximum.

To gain a qualitative understanding of what sort of pairings can induce an entropy minimum at $\varepsilon_{\text{VH}}$ strain, it is sufficient to consider the $\gamma$ band near the Van Hove lines.
This is justified by the fact that the $\gamma$ band contributes \SI{60}{\percent} of the total DOS [Appendix~\ref{sec:SRO-TBA}] and is solely responsible for the normal-state peak in the entropy.
For the moment, we shall also neglect the $k_3$-dispersion.

The DOS of a band in 2D with a dispersion $\epsilon_{\vb{k}}$ and SC gap $\Delta_{\vb{k}}$ is given by:
\begin{equation}
g_{\text{sc}}(E) = 2 \int \frac{\dd{k_1} \dd{k_2}}{(2 \pi)^2} \Dd\!\big(E - \xi_{\vb{k}}\big),
\end{equation}
where the $2$ is due to spin and $\xi_{\vb{k}} = \sqrt{\epsilon_{\vb{k}}^2 + \abs{\Delta_{\vb{k}}}^2}$ is the Bogoliubov quasi-particle dispersion.
It is often easier to calculate the integrated DOS
\begin{equation}
\mathcal{N}_{\text{sc}}(E) = \int_0^E \dd{E'} \, g_{\text{sc}}(E') = 2 \int_{\xi_{\vb{k}} \leq E} \frac{\dd{k_1} \dd{k_2}}{(2 \pi)^2}
\end{equation}
and then differentiate it to get $g_{\text{sc}}(E)$.
Near the Van Hove point $\left(0, \pi\right)$, the dispersion of the $\gamma$ band is approximately given by [see Appendix~\ref{sec:SRO-TBA} or Eq.~\eqref{eq:VHdispexpansion}]:
\begin{equation}
\epsilon_{\vb{k}} = \frac{1}{2m_1} k_1^2 - \frac{1}{2m_2} k_2^2 = \frac{1}{m_{*}} q_{+} q_{-}, \label{eq:saddle-epsilon}
\end{equation}
where $m_{*} = \sqrt{m_1 m_2} = 1/\SI{3200}{\kelvin}$, $r = \sqrt[4]{m_2 / m_1} = 0.59$, and $q_{\pm} = \frac{1}{\sqrt{2}} \left(r k_1 \pm k_2 / r\right)$.
Since this expression for $\epsilon_{\vb{k}}$ only applies near the Van Hove point, we impose a momentum cutoff $\abs{q_{\pm}} \leq \Lambda$.

In the normal state (NS), $\Delta_{\vb{k}} = 0$ and the DOS at the Van Hove strain equals:
\begin{equation}
g_{\text{sc}}^{\text{NS}}(E) = \frac{8 m_{*}}{(2 \pi)^2} \log\frac{\Lambda^2}{m_{*} E}. \label{eq:VH-DOS-NS}
\end{equation}
This diverges logarithmically as $E \to 0$.
As we move away from $\varepsilon_{100} = \varepsilon_{\text{VH}}$, the logarithmic divergence is moved away from the Fermi level $E = 0$, explaining the normal-state entropy maximum.

If we fully gap (FG) the saddle point, $\Delta_{\vb{k}} = \Delta_0$, then the DOS vanishes up to $\Delta_0$, $g_{\text{sc}}^{\text{FG}}(E \leq \Delta_0) = 0$, and diverges above it according to ($E > \Delta_0$):
\begin{equation}
g_{\text{sc}}^{\text{FG}}(E) = \frac{8 m_{*}}{(2 \pi)^2} \frac{E}{\sqrt{E^2 - \Delta_0^2}} \log\frac{\Lambda^2}{m_{*} \sqrt{E^2 - \Delta_0^2}}. \label{eq:VH-DOS-FG}
\end{equation}
Since $\Dd_{T}(E)$ in Eq.~\eqref{eq:Fermi-liquid-entropy} has a width $\sim k_B T$, for sufficiently large $\Delta_0 / k_B T$ the normal-state entropy maximum can be suppressed so strongly that it becomes a minimum as a function of strain.
Hence fully gapping the Van Hove lines reproduces the features of Figure~\ref{fig:elasto}.
Note that a constant gap does not necessarily mean an $s$-wave state, but merely that the gap is finite in the vicinity of the Van Hove point.
For instance, $d_{x^2-y^2}$-wave pairing is finite at the Van Hove point $\left(0, \pi\right)$ and approximately constant around it.
Our analysis focuses only on the behavior of the pairing gap near the saddle point of the dispersion.

Can pairings with nodal lines at the Van Hove lines also reproduce the SC entropy minimum?
To answer this question, let us calculate the DOS for a vertical and horizontal line node.
For vertical line nodes (VLN), there are two cases to distinguish: when $\Delta_{\vb{k}}$ is linear and when $\Delta_{\vb{k}}$ is quadratic in $\vb{k}$.

In the linear case, we may always write the gap as:
\begin{equation}
\Delta_{\vb{k}} = \Delta_0 \left(q_{+} \cos \varphi + q_{-} \sin \varphi\right) / \Lambda = \Delta_0 (p_1 / \Lambda).
\end{equation}
In the limit of small $E$, the inequality $\xi_{\vb{k}} \leq E$ that determines $\mathcal{N}_{\text{sc}}(E)$ simplifies to
\begin{equation}
\frac{\Delta_0^2}{\Lambda^2} p_1^2 + \frac{\sin^2(2 \varphi)}{4 m_{*}^2} p_2^4 \leq E^2,
\end{equation}
where $p_2 = q_{-} \cos \varphi - q_{+} \sin \varphi$.
The area enclosed by this inequality equals $\pi' \abs{p_{1, \text{max}}}_E \abs{p_{2, \text{max}}}_E$, where $\pi' = 4 \int_0^1 \dd{x} \sqrt{1-x^4} \approx 3.496$, and therefore for small $E$:
\begin{equation}
g_{\text{sc}}^{\text{VLN}}(E \to 0) = \frac{3 \pi'}{(2 \pi)^2} \frac{\Lambda}{\Delta_0} \sqrt{\frac{2 m_{*} E}{\abs{\sin 2 \varphi}}}. \label{eq:VH-DOS-VLN}
\end{equation}
This $g_{\text{sc}}^{\text{VLN}} \propto \sqrt{E}$ behavior persists up to the point where $g_{\text{sc}}^{\text{VLN}}(E_w) \approx g_{\text{sc}}^{\text{NS}}(E_w)$.
By solving this equation with $\Delta_0 \sim \SI{3}{\kelvin}$ (the $T_c$ at $\varepsilon_{xx} = \varepsilon_{\text{VH}}$) and $\Lambda \sim 0.5$, one obtains $E_w \sim \SI{0.2}{\kelvin}$.\textsuperscript{\citenum{Note9}}
Exceptionally, when $\varphi = 0$ or $\pi/2$, one finds a constant DOS up to $\Delta_0$:
\begin{equation}
g_{\text{sc}}^{\text{VLN}'}(E \leq \Delta_0) = \frac{8 m_{*}}{(2 \pi)^2} \arcsinh\frac{\Lambda^2}{m_{*} \Delta_0}.
\end{equation}
Thus if a single line node cuts through the Van Hove point, the DOS generically vanishes like $\sqrt{E}$ in a very narrow range $E \lesssim \SI{0.2}{\kelvin}$.
If this line node is fine-tuned to coincide with the lines $q_{+} = 0$ or $q_{-} = 0$, then the DOS becomes finite and large.

\footnotetext[9]{ 
The solution of $\sqrt{x} = \frac{1}{2} \delta \log(1/x)$ is $x = \delta^2 W^2(1/\delta)$, where $W(x)$ is the Lambert $W$-function. In our case $x = m_{*} E_w / \Lambda^2$ and $\delta = (8 \sqrt{2} / 3 \pi') (m_{*} \Delta_0 / \Lambda^2)$.
}

The second case is when $\Delta_{\vb{k}}$ is quadratic in $\vb{k}$.
Quadratic $\Delta_{\vb{k}}$ may correspond to a line node with a quadratic orthogonal dispersion, a pair of line nodes that intersect at $\vb{k} = \vb{0}$, or a point node, depending on the eigenvalues of the Hessian.
The inequality $\xi_{\vb{k}} \leq E$ is in this case invariant under the scaling $\vb{k} \mapsto \sqrt{\alpha} \vb{k}$, $E \mapsto \alpha E$.
Hence $\mathcal{N}_{\text{sc}}(E)$ is linear in $E$ for small $E$, yielding a finite $g_{\text{sc}}^{\text{VLN}''}(E = 0)$ and no opening of a gap.
Exceptionally, when we have two SC line nodes that coincide with the Van Hove strain Fermi surfaces $q_{\pm} = 0$, the SC gap equals $\Delta_{\vb{k}} = \Delta_0 (q_{+} q_{-} / \Lambda^2)$, from which we see that $g_{\text{sc}}^{\text{VLN}'''}$ retains the normal-state logarithmic singularity, albeit with a renormalized $1/m_{*} \mapsto \sqrt{1/m_{*}^2 + \Delta_0^2 / \Lambda^4}$.

\begin{figure}[t]
\includegraphics[width=\columnwidth]{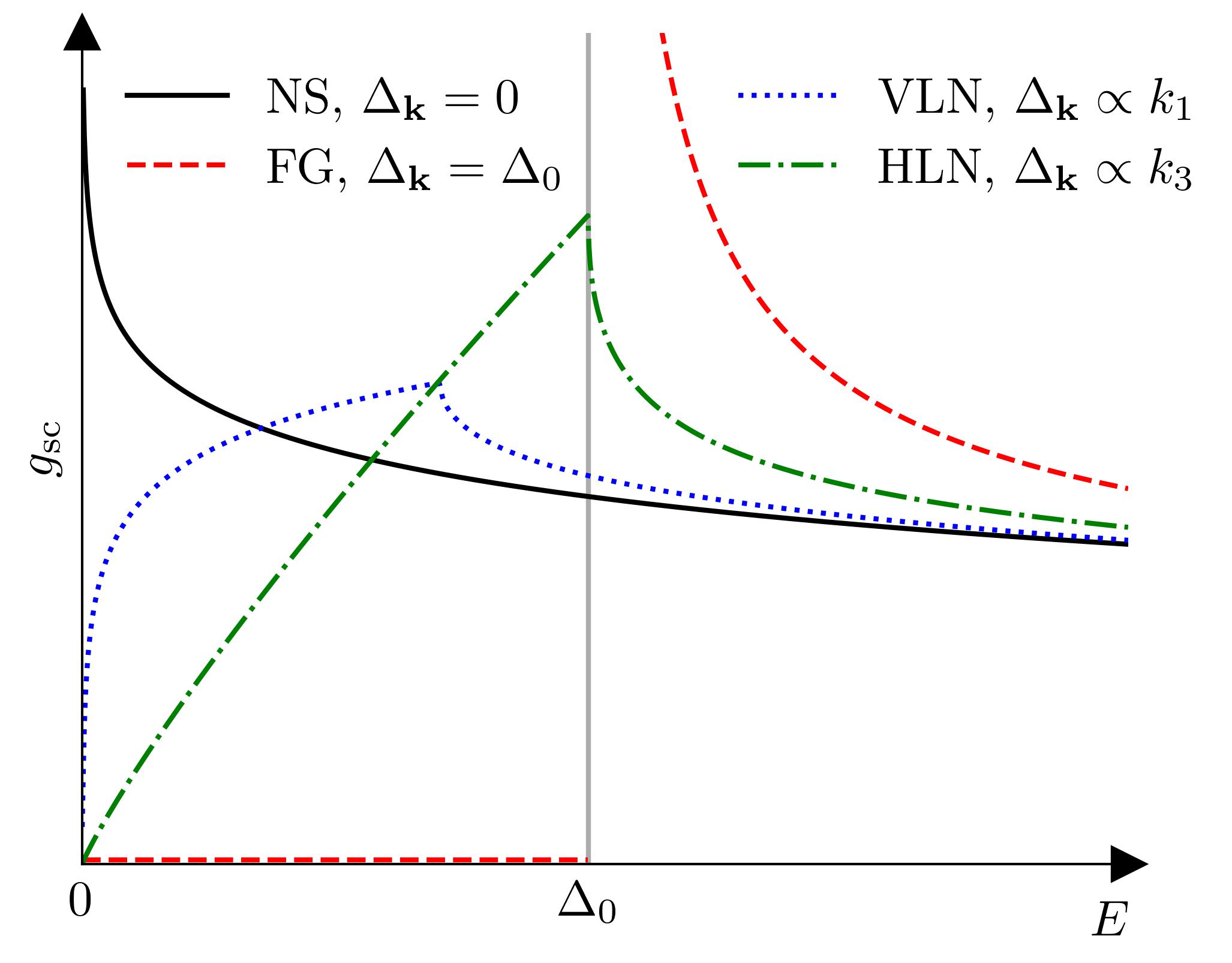}
\caption{The Van Hove line contributions to the DOS $g_{\text{sc}}(E)$ for the four possible types of gaps considered in the text: normal state (NS), SC with a fully gapped (FG) Van Hove line, SC with a vertical line node (VLN) on the Van Hove line, and SC with a horizontal line node (HLN) crossing the Van Hove line.
These correspond to Eqs.~\eqref{eq:VH-DOS-NS}, \eqref{eq:VH-DOS-FG}, \eqref{eq:VH-DOS-VLN}, and \eqref{eq:VH-DOS-HLN}, respectively.
The VLN case (with $\varphi = \pi/4$) was calculated numerically.
The parameter values $m_{*}^{-1} = \SI{3200}{\kelvin}$, $\Delta_0 = \SI{3}{\kelvin}$, and $\Lambda = 0.5$ were used in all four cases.
Note that the Fermi energy ($E = 0$) is tuned precisely to the saddle point, so this depicts the DOS at the Van Hove strain $\varepsilon_{100} = \SI{-0.44}{\percent}$.}
\label{fig:DOS}
\end{figure}

Lastly, there's the possibility of a horizontal line node (HLN) crossing the vertical Van Hove line $\left(0, \pi, k_3\right)$.
For a schematic $\Delta_{\vb{k}} = \Delta_0 (k_3 / \pi)$, the 3D DOS can be calculated by averaging Eq.~\eqref{eq:VH-DOS-FG}:
\begin{equation}
\begin{aligned}
g_{\text{sc}}^{\text{HLN}}(E) &= \int_{- \pi}^{\pi} \frac{\dd{k_3}}{2 \pi} \left.g_{\text{sc}}^{\text{FG}}(E)\right|_{\Delta_0 \to \Delta_0 \abs{k_3} / \pi} \\
&= \frac{4 m_{*}}{(2 \pi)^2} \frac{E}{\Delta_0} \left[\pi \log\frac{2 \Lambda^2}{m_{*} E} - \zeta(E)\right],
\end{aligned} \label{eq:VH-DOS-HLN}
\end{equation}
where $\zeta(E \leq \Delta_0) = 0$ and for $E > \Delta_0$:
\begin{align}
\zeta(E) &= (\pi - 2 \arccos x) \log\frac{\Lambda^2}{m_{*} E} + 2 \arcsin x \log(2 x) \notag \\
&\quad - 2 \arctan\frac{x}{\sqrt{1-x^2}} \log x + \operatorname{Cl}_2(\phi),
\end{align}
where $x = \sqrt{1 - \Delta_0^2/E^2}$, $\phi = \arccos(1 - 2 x^2)$, and $\operatorname{Cl}_2(\phi) = \sum_{k=1}^{\infty} \sin(k \phi)/k^2$ is the Clausen function.
$g_{\text{sc}}^{\text{HLN}}$ is thus roughly linear in $E$ up to $\Delta_0$.

The dependence of the DOS $g_{\text{sc}}(E)$ for different realizations of the SC gap $\Delta_{\vb{k}}$ near the saddle point is summarized in Figure~\ref{fig:DOS}.

Now we come back to the question of whether line nodes at the Van Hove lines are consistent with an entropy minimum.
To clarify this issue, we need to take into account the $k_3$-dispersion, the energy integral in Eq.~\eqref{eq:Fermi-liquid-entropy}, and the DOS contributions of the other bands.

The $k_3$-dispersion of the $\gamma$ band smears all characteristically 2D features of the DOS by the scale of its energy variation $\var{\epsilon_{\text{VH}}} \sim \SI{2}{\kelvin}$ [Eq.~\eqref{eq:VHdispexpansion}].
The normal-state logarithmic singularity becomes a peak.
The $g_{\text{sc}}^{\text{VLN}} \propto \sqrt{E}$ ascent is cut off to give a finite zero-energy DOS that is because of $E_w / \var{\epsilon_{\text{VH}}} \ll 1$ of the same magnitude as the normal-state DOS.
Finally, the HLN DOS attains a finite zero-energy DOS that is at most a factor of three or so smaller than the normal-state DOS (since $\var{\epsilon_{\text{VH}}} / \Delta_0 \sim 1$).
The $\Dd_T(E)$ factor in Eq.~\eqref{eq:Fermi-liquid-entropy} leads to a temperature smearing that has a similar effect: the ``effective DOS'' that enters the entropy is not $g_{\text{sc}}(0)$, but $g_{\text{sc}}(E)$ averaged over $E \sim k_B T$.
All in all, because of these smearing effects, vertical line nodes at the Van Hove lines $\left(0, \pm\pi, k_3\right)$ do not suppress the entropy contribution coming from the Van Hove lines, whereas horizontal line nodes can indeed suppress it.

Because of the strain-dependence of $T_c$, the SC gap becomes $\varepsilon_{100}$-dependent at constant $T$, peaking at Van Hove strain.
A strong enough gapping of the $\alpha$ and $\beta$ bands could then, in principle, suppress the entropy more than the Van Hove singularities enhance it, resulting in a minimum.
To exclude this scenario, we have calculated the entropy for when the $\alpha$, $\beta$, and \SI{80}{\percent} of the $\gamma$ band have $\Delta_{\vb{k}} = \Delta_0$, and the remaining \SI{20}{\percent} of the $\gamma$ band that includes the Van Hove lines has $\Delta_{\vb{k}} = 0$.\textsuperscript{\citenum{Note10}}
The result of this calculation is that a minimum as a function of strain does develop, but the drop in the entropy is \SI{20}{\percent} too small when compared to experiment at $\SI{2.5}{\kelvin}$.
Thus even in this worst-case scenario, where line nodes that are known~\cite{NishiZaki2000, Deguchi2004, Kittaka2018, Matsui2001, Lupien2001, Ishida2000, Bonalde2000, Deguchi2004-p2} to be present in the system are neglected, the Van Hove lines must be gapped in some way to agree with experiment.

\footnotetext[10]{ 
For the total DOS and gap we have assumed the form:
\begin{align*}
g_{\text{sc}}^{\text{tot}}(E) &= g_{\text{VH}} + \Theta(E - \Delta_0) \frac{E}{\sqrt{E^2 - \Delta_0^2}} g_{\text{rest}}, \\
\Delta_0 &= 1.76 \, k_B T_c \tanh\!\bigg(1.76 \sqrt{\frac{T_c}{T} - 1}\bigg),
\end{align*}
where $g_{\text{VH}}$ is the normal-state DOS coming from the parts of the $\gamma$ sheet that are close to the Van Hove lines and $g_{\text{rest}}$ is the remaining normal-state DOS.
Both $g_{\text{VH}}(\varepsilon_{100}) + g_{\text{rest}}(\varepsilon_{100}) \propto \left.S(\varepsilon_{100}, T) / T\right|_{T > T_c}$ and $T_c(\varepsilon_{100})$ are known experimentally.
Only the ratio $g_{\text{VH}} / g_{\text{rest}}$ needs to be calculated, which we have done using the tight-bind model of Appendix~\ref{sec:SRO-TBA}.
One finds that $g_{\text{rest}}(\varepsilon_{100})$ is roughly strain-independent, as expected.
The entropy is given by Eq.~\eqref{eq:Fermi-liquid-entropy}.
}

The final conclusion that follows from all of these considerations is that the Van Hove lines $\vb{k}_{\text{VH}} \approx \left(0, \pm \frac{\pi}{a}, k_3\right)$ must be either fully gapped or can at most have a horizontal line node crossing them.
Hence, we may exclude vertical line nodes at $\vb{k}_{\text{VH}}$ near Van Hove strain~\cite{Li2022}.
That the heat capacity jump is maximal at the Van Hove strain~\cite{Li2021} also supports this conclusion.
Vertical line nodes away from the Van Hove lines are still possible.

To draw conclusions for the unstrained tetragonal system from measurements performed at uniaxial strain $\varepsilon_{100} \approx \varepsilon_{\text{VH}}$, we rely on the assumption that the pairing states of the strained and unstrained system are adiabatically connected.
Measurements of the highly-sensitive elastocaloric effect~\cite{Li2022} and heat capacity~\cite{Li2021} show no hints of a transition between two different bulk SC states under $[100]$ strain.
By contrast, the onset of spin-density waves, previously found through muon spin relaxation~\cite{Grinenko2021-unaxial}, is clearly visible in the elastocaloric data of Ref.~\cite{Li2022}.
So the elastocaloric effect is able to identify a variety of phase transitions.

We may thus exclude all SC states of the unstrained system that are adiabatically connected to SC states of the $\varepsilon_{xx}$ strained system which have a vertical line node at $\vb{k}_{\text{VH}} \approx \left(0, \pm \frac{\pi}{a}, k_3\right)$.
Given that $\varepsilon_{xx}$ strain preserves all the symmetry operations that map the Van Hove lines to themselves, as we shall see in Sec.~\ref{sec:behavior-vH}, we may conclude that there are no vertical line nodes at either $\left(\pm \frac{\pi}{a}, 0, k_3\right)$ nor $\left(0, \pm \frac{\pi}{a}, k_3\right)$ in the unstrained tetragonal system.
Intuitively, this means that SRO's SC takes full advantage of the enhanced DOS induced by the Van Hove lines.
Indeed, the drastic enhancement of $T_c$ and $B_{c2}$ under uniaxial pressure~\cite{Hicks2014, Taniguchi2015, Steppke2017, Barber2019, Jerzembeck2023} were suggestive of this conclusion long ago, but only with the recent elastocaloric measurements of Ref.~\cite{Li2022} could more conclusive statements be made.

\section{Location of the Van Hove lines} \label{sec:vH-line}
Here we establish that the Van Hove lines are adequately approximated with $\left(\pm \frac{\pi}{a}, 0, k_3\right)$ and $\left(0, \pm \frac{\pi}{a}, k_3\right)$.
For a simple-tetragonal lattice, the Van Hove lines are lines of high symmetry.
However, they are not located precisely on the boundary of the body-centered first Brillouin zone relevant here, which could in principle allow for large deviations away from $\left(\pm \frac{\pi}{a}, 0, k_3\right)$ and $\left(0, \pm \frac{\pi}{a}, k_3\right)$.
As we shall see, the high anisotropy of SRO makes these deviations negligible, justifying the subsequent analysis.

Van Hove points are points in momentum space where the gradient of the band energy $\epsilon_{\vb{k}}$ vanishes.
In 3D, the solutions of $\grad \epsilon_{\vb{k}} = \vb{0}$ are generically isolated points.
However, quasi-2D dispersions may yield Van Hove \emph{lines}, that is, lines on which a number of Van Hove points are situated of similar energy.
The quality of the emergent Van Hove lines is quantified by how well-aligned the Van Hove points are to a line and by how close the energies of the Van Hove points are.

Consider the Van Hove line $\left(0, \frac{\pi}{a}, k_3\right)$.
Then for any two $\vb{k} = \left(\var{k_1}, \frac{\pi}{a} + \var{k_2}, k_3\right)$ and $\vb{k}' = R(g) \vb{k}$ related by a symmetry operation $g \in D_{4h}$, $\epsilon_{\vb{k}} = \epsilon_{\vb{k}' + \vb{K}}$ for any reciprocal lattice vector $\vb{K}$.
Applying this to parity gives $\grad \epsilon_{\vb{k}} = \vb{0}$ at the mid-points of the Brillouin zone faces, which for body-centered tetragonal SRO are $\left(0, \frac{\pi}{a}, \pm \frac{\pi}{c}\right)$.
These are the first two Van Hove points.
The positions of the other two Van Hove points are restricted by symmetry to be at $\left(0, \frac{\pi}{a} + \var{k_{\text{VH}, 2}}, 0\right)$ and $\left(0, \frac{\pi}{a} - \var{k_{\text{VH}, 2}}, \pm \frac{2 \pi}{c}\right)$.
Reflection across the $k_1 = 0$ plane implies $\partial_{k_1} \epsilon_{\vb{k}} = 0$ in the $k_1 = 0$ plane and reflection across the $k_3 = 0$ plane implies $\partial_{k_3} \epsilon_{\vb{k}} = 0$ in the planes $k_3 = 0, \pm \frac{2 \pi}{c}$.
If the system were simple tetragonal-periodic, then reflection across the $k_2 = 0$ plane would imply $\partial_{k_2} \epsilon_{\vb{k}} = 0$ in the $k_2 = \pm \frac{\pi}{a}$ planes, making $\var{k_{\text{VH}, 2}} = 0$.
Because of the smallness of the characteristically body-centered hopping in SRO, which is always between layers, $\var{k_{\text{VH}, 2}}$ is very close to zero.

\begin{figure}[t]
\includegraphics[width=\columnwidth]{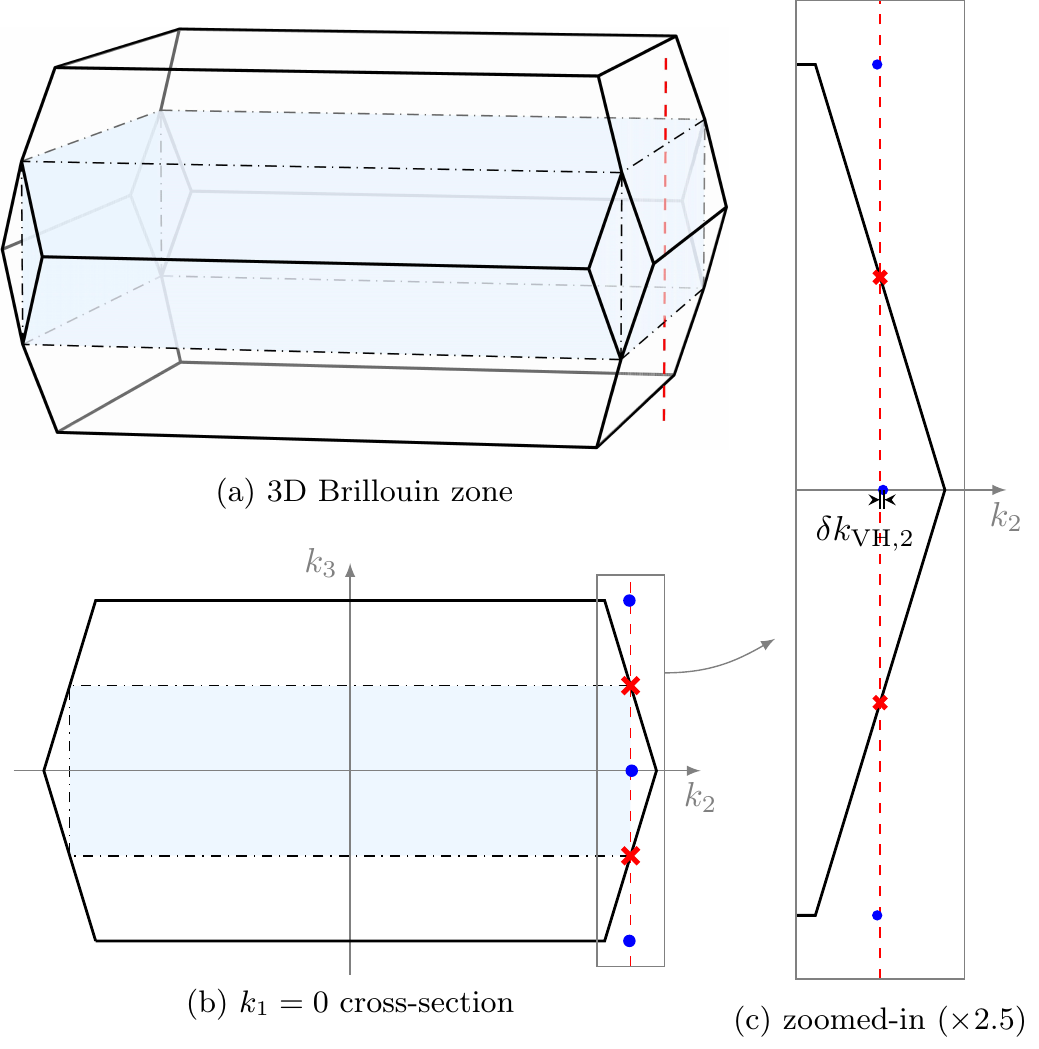}
\caption{The body-centered tetragonal Brillouin zone of SRO~(a), its $k_1 = 0$ cross-section~(b), and the region around the $\left(0, \frac{\pi}{a}, k_3\right)$ Van Hove line~(c).
Shaded in blue is the simple tetragonal Brillouin zone.
The red crosses are the $\left(0, \frac{\pi}{a}, \pm \frac{\pi}{c}\right)$ Van Hove points.
The blue dots are the $\left(0, \frac{\pi}{a} + \var{k_{\text{VH}, 2}}, 0\right)$ and $\left(0, \frac{\pi}{a} - \var{k_{\text{VH}, 2}}, \pm \frac{2 \pi}{c}\right)$ Van Hove points.
Together they constitute the Van Hove line $\left(0, \frac{\pi}{a}, k_3\right)$, drawn here with a dashed red line.
The displacement length $\var{k_{\text{VH}, 2}} \approx 0.013 / a$ is designated in~(c).}
\label{fig:vanHove}
\end{figure}

From the tight-binding model [Appendix~\ref{sec:SRO-TBA}], we may extract the following simplified expression for the dispersion of the $\gamma$ band near the Van Hove line $\left(0, \frac{\pi}{a}, k_3\right)$:
\begin{align}
\begin{aligned}
\epsilon_{\vb{k}} &= \mu_{\text{VH}} + \frac{a^2}{2 m_1} k_1^2 - \frac{a^2}{2 m_2} \left(k_2 - \frac{\pi}{a}\right)^2 \\
&- \var{\epsilon_{\text{VH}}} \cos c k_3 + \frac{a^2}{m_2} \var{k_{\text{VH}, 2}} \left(k_2 - \frac{\pi}{a}\right) \cos\frac{c k_3}{2}.
\end{aligned}  \label{eq:VHdispexpansion}
\end{align}
Its form follows from symmetry; only the lowest powers in $k_1, k_2$ and lowest harmonics in $k_3$ were retained.
Here $\mu_{\text{VH}} = \SI{54}{\milli\electronvolt}$, $\var{\epsilon_{\text{VH}}} = \SI{2.4}{\kelvin}$, $\var{k_{\text{VH}, 2}} = 0.013/a$, $m_1^{-1} = \SI{1100}{\kelvin}$, and $m_2^{-1} = \SI{9300}{\kelvin}$.
While this dispersion was derived from a model of unstrained SRO, it offers a good understanding of the effects of the $k_3$-dispersion on the Van Hove line.
The deviation of the Van Hove points from the $(\frac{\pi}{a}, 0, k_3)$-line is characterized by $\var{k_{\text{VH}, 2}} \ll \frac{2 \pi}{a}$, which is a factor of $500$ smaller than the width of the Brillouin zone.
Furthermore, the difference in the $\gamma$ band energies of the Van Hove points is given by $\var{\epsilon_{\text{VH}}}$ which is on the order of a few kelvins.
We may thus conclude that the four Van Hove points, illustrated in Figure~\ref{fig:vanHove}, together constitute a Van Hove line $\left(0, \frac{\pi}{a}, k_3\right)$ to a high degree of accuracy.
The same is true for the Van Hove lines $\left(0, - \frac{\pi}{a}, k_3\right)$ and $\left(\pm \frac{\pi}{a}, 0, k_3\right)$.

\section{Behavior on the Van Hove lines} \label{sec:behavior-vH}
To see which SC states are excluded by the fact that vertical line nodes on the Van Hove lines are incompatible with the elastocaloric effect data, we first need to see which SC states are possible.
This is significant because the multiband nature of SRO allows for a richer set of possibilities than usual.
Since this has already been analyzed~\cite{Ramires2019, Kaba2019, *Kaba2019-E, Huang2019}, here we only briefly discuss how the multiband case differs from the singleband one, delegating the details of the categorization of all possible SC states to Appendix~\ref{sec:SC-construct}.

To describe SRO's SC, we employ an effective model based on the $t_{2g}$ orbitals of \ce{Ru} [Appendix~\ref{sec:SRO-TBA}].
Within it, SC is described by a gap matrix $\Delta_{\alpha \beta}(\vb{k})$ which is characterized by its momentum dependence and spin-orbit structure.
It is the possibility of a non-trivial orbital structure that sets multiband systems apart from singleband ones.
Thus, for instance, when dealing with even pairings, we cannot simply assume a spin singlet that transforms trivially ($A_{1g}$) under all symmetry operations and equate the irrep of the momentum wavefunction with the irrep of the total gap matrix.
The irrep of the gap matrix is determined by the \emph{product} of the irreps of its momentum and spin-orbit parts.
Within the effective model, there are spin-orbit matrices belonging to all the possible irreps of $D_{4h}$ for both even and odd pairings.
The details of how $\Delta_{\alpha \beta}(\vb{k})$ are constructed by combining pairing wavefunctions $d(\vb{k})$ with spin-orbit matrices $\Gamma$ can be found in Appendix~\ref{sec:SC-construct}.

Now we analyze which SC states of the $\varepsilon_{xx}$ strained system gap the Van Hove lines sufficiently strongly to be able to explain the elastocaloric experiment~\cite{Li2022}.
Viable unstrained SC states must be adiabatically connected to these states.
As we shall see, in the arguments of this section the key symmetry operations are those that map the Van Hove lines $\vb{k}_{\text{VH}} = \left(0, \pm \frac{\pi}{a}, k_3\right)$ to themselves.
As it turns out, although $\varepsilon_{xx}$ strain reduces the point group from $D_{4h}$ to $D_{2h}$ [Table~\ref{tab:D2h-char}], the symmetries that map the Van Hove lines to themselves are the same for both $D_{4h}$ and $D_{2h}$.
Hence we may do the whole analysis either with or without $\varepsilon_{xx}$ strain.
We have opted for the latter.
Using Table~\ref{tab:D4h-to-D2h}, one may translate all the results for irreps of $D_{4h}$ of this section into results for irreps of $D_{2h}$.
Table~\ref{tab:D4h-to-D2h} also specifies which irreps of $D_{2h}$ are adiabatically connected to which irreps of $D_{4h}$, which brings us back to the initial $D_{4h}$ irreps.

\begin{table}[t]
\caption{The character table of the point group $D_{2h}$~\cite{Dresselhaus2007}.
Irreps are divided into even ($g$) and odd ($u$) ones.
Primes have been added on the irreps to distinguish them from $D_{4h}$ irreps.
$C_2^x$, $C_2^y$, and $C_2^z$ are rotations by $\pi$ around the $x$, $y$, and $z$ axes, respectively.
$P$ is parity.
$\Sigma_x$, $\Sigma_y$, $\Sigma_z$ are compositions of $C_2^x$, $C_2^y$, and $C_2^z$ with $P$, respectively.}
{\renewcommand{\arraystretch}{1.3}
\renewcommand{\tabcolsep}{7.6pt}
\begin{tabular}{c|rrrr|rrrr} \hline\hline
$D_{2h}$ & $E$ & $C_2^z$ & $C_2^y$ & $C_2^x$ & $P$ & $\Sigma_z$ & $\Sigma_y$ & $\Sigma_x$
\\ \hline
$A_{1g}'$ & $1$ & $1$ & $1$ & $1$ & $1$ & $1$ & $1$ & $1$
\\
$B_{1g}'$ & $1$ & $1$ & $-1$ & $-1$ & $1$ & $1$ & $-1$ & $-1$
\\
$B_{2g}'$ & $1$ & $-1$ & $1$ & $-1$ & $1$ & $-1$ & $1$ & $-1$
\\
$B_{3g}'$ & $1$ & $-1$ & $-1$ & $1$ & $1$ & $-1$ & $-1$ & $1$
\\ \hline
$A_{1u}'$ & $1$ & $1$ & $1$ & $1$ & $-1$ & $-1$ & $-1$ & $-1$
\\
$B_{1u}'$ & $1$ & $1$ & $-1$ & $-1$ & $-1$ & $-1$ & $1$ & $1$
\\
$B_{2u}'$ & $1$ & $-1$ & $1$ & $-1$ & $-1$ & $1$ & $-1$ & $1$
\\
$B_{3u}'$ & $1$ & $-1$ & $-1$ & $1$ & $-1$ & $1$ & $1$ & $-1$
\\ \hline\hline
\end{tabular}}
\label{tab:D2h-char}
\end{table}

\begin{table}[t]
\caption{Reduction of the $D_{4h}$ irreps (top) to $D_{2h}$ irreps (bottom) that occurs under $\varepsilon_{xx}$ uniaxial strain.
Parity stays the same so we have suppressed the $g/u$ subscripts.
$\{x \mid y\}$ transforms according to the $\rho^{(E)}(g)$ of Eq.~\eqref{eq:E-rep-rho}, Appendix~\ref{sec:SC-construct}.}
\includegraphics[width=\columnwidth]{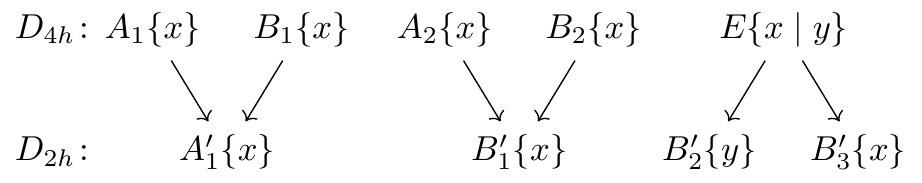}
\label{tab:D4h-to-D2h}
\end{table}

Let us consider the Van Hove line $\vb{k}_{\text{VH}} = \left(0, \frac{\pi}{a}, k_3\right)$.
For a SC gap matrix $\Delta_a(\vb{k})$ to be able to gap the $\gamma$ band at $\vb{k}_{\text{VH}}$, both its pairing wavefunction $d_a(\vb{k})$ and the projection of its spin-orbit matrix $\Gamma_a$ onto the $\gamma$ band must be finite there.

The only point group symmetries $g \in D_{4h}$ that constrain $d_a\!\left(\vb{k}_{\text{VH}}\right)$ or the band projections of $\Gamma_a$ are those that map the $\left(0, \frac{\pi}{a}, k_3\right)$ line to itself, modulo body-centered reciprocal lattice vectors.
One readily find that these are
\begin{align}
\begin{aligned}
\Sigma_x'\colon k_3 &\mapsto k_3, \\
\Sigma_y', C_{2z}\colon k_3 &\mapsto k_3 + \frac{2 \pi}{c}, \\
\Sigma_h, C_{2y}'\colon k_3 &\mapsto - k_3, \\
C_{2x}'\colon k_3 &\mapsto - k_3 + \frac{2 \pi}{c}.
\end{aligned}
\end{align}
Here, $C_{2x}'$, $C_{2y}'$, $C_{2z}$ are rotations by $\pi$ around $x$, $y$, and $z$, respectively, and $\Sigma_x' = P C_{2x}'$, $\Sigma_y' = P C_{2y}'$, $\Sigma_h = P C_{2z}$ are reflections.
Given that $C_{2z} = \Sigma_x' \Sigma_y'$ and $C_{2y}' = \Sigma_x' \Sigma_h$, we may focus solely on the reflections and $C_{2x}'$.
Their matrices are listed in Table~\ref{tab:rho-kvH}.
The strongest constraints follow from $\Sigma_x'$ because it maps $k_3 \mapsto k_3$.
In the simple tetragonal limit, $k_3 \cong k_3 + \frac{2 \pi}{c}$ so $\vb{k}_{\text{VH}}$ are on the Brillouin zone boundary and $\Sigma_y', C_{2z}$ give strong constraints too.

\begin{table}[t]
\caption{The $\rho_{ab}^{(\lambda)}(g)$ of non-trivial irreps $\lambda$ of $D_{4h}$.}
{\renewcommand{\arraystretch}{1.3}
\renewcommand{\tabcolsep}{2.0pt}
\begin{tabular}{c|ccccccccc} \hline\hline
$g$
& $A_{2g}$ & $B_{1g}$ & $B_{2g}$ & $E_{g}$
& $A_{1u}$ & $A_{2u}$ & $B_{1u}$ & $B_{2u}$ & $E_{u}$
\\ \hline
$\Sigma_x'$
& $-1$ & $1$ & $-1$ & $\begin{pmatrix} 1 & 0 \\ 0 & -1 \end{pmatrix}$
& $-1$ & $1$ & $-1$ & $1$ & $\begin{pmatrix} -1 & 0 \\ 0 & 1 \end{pmatrix}$
\\
$\Sigma_y'$
& $-1$ & $1$ & $-1$ & $\begin{pmatrix} -1 & 0 \\ 0 & 1 \end{pmatrix}$
& $-1$ & $1$ & $-1$ & $1$ & $\begin{pmatrix} 1 & 0 \\ 0 & -1 \end{pmatrix}$
\\
$\Sigma_h$
& $1$ & $1$ & $1$ & $\begin{pmatrix} -1 & 0 \\ 0 & -1 \end{pmatrix}$
& $-1$ & $-1$ & $-1$ & $-1$ & $\begin{pmatrix} 1 & 0 \\ 0 & 1 \end{pmatrix}$
\\
$C_{2x}'$
& $-1$ & $1$ & $-1$ & $\begin{pmatrix} 1 & 0 \\ 0 & -1 \end{pmatrix}$
& $1$ & $-1$ & $1$ & $-1$ & $\begin{pmatrix} 1 & 0 \\ 0 & -1 \end{pmatrix}$
\\ \hline\hline
\end{tabular}}
\label{tab:rho-kvH}
\end{table}

Consider one of the four $g$ from Table~\ref{tab:rho-kvH} and a $k_3$ that $g$ maps to itself, modulo $\frac{4 \pi}{c}$.
Periodicity and the symmetry transformation rule of pairing wavefunctions [Eq.~\eqref{eq:transf-d}, Appendix~\ref{sec:SC-construct}] then give the constraint:
\begin{align}
d_a\!\left(0, \tfrac{\pi}{a}, k_3\right) &= \sum_{b = 1}^{\dim \lambda} \rho_{ab}^{(\lambda)}(g) d_b\!\left(0, \tfrac{\pi}{a}, k_3\right).
\end{align}
By analyzing it, we find the following symmetry-enforced behavior of $d_a\!\left(0, \frac{\pi}{a}, k_3\right)$, depending on its irrep and $k_3$:
\begin{itemize}
\item $d$ belonging to $A_{2g}$, $B_{2g}$, $A_{1u}$, and $B_{1u}$ vanish for all $k_3$.
\item For $\{d_1 \mid d_2\} \in E_g$, $d_2$ vanishes for all $k_3$, whereas $d_1$ vanishes only at $k_3 = 0, \pm \frac{2 \pi}{c}$.
\item For $\{d_1 \mid d_2\} \in E_u$, $d_1$ vanishes for all $k_3$, whereas $d_2$ vanishes only at $k_3 = \pm \frac{\pi}{c}$.
\item For those $\{d_1 \mid d_2\} \in E_{g/u}$ that are periodic under simple tetragonal translations [$k_3 \cong k_3 + \frac{2 \pi}{c}$], both components vanish for all $k_3$.
\item $d$ from irreps $A_{2u}$ and $B_{2u}$ vanish only at $k_3 = 0$, $\pm \frac{\pi}{c}$, and $\pm \frac{2 \pi}{c}$, but are otherwise unconstrained.
\item $d$ from $A_{1g}$ and $B_{1g}$ are completely unconstrained for all $k_3$.
\end{itemize}

To proceed, we consider the pairing of the band eigenstates of the problem and focus on intraband pairing.
To find it, we need to project $\Gamma_a$ onto the bands.
Call $V_{\vb{k}} = \left(v_{\vb{k} \uparrow}, v_{\vb{k} \downarrow}\right)$ the Kramers-degenerate eigenvectors of the $\gamma$ band, $H_{\vb{k}} V_{\vb{k}} = \epsilon_{\vb{k}} V_{\vb{k}}$.
The projection is then given by:
\begin{align}
P_{\vb{k} a} &= V_{\vb{k}}^{\dag} \Gamma_a V_{- \vb{k}}^{*} = \sum\nolimits_{\mu} P_{a}^{\mu}(\vb{k}) \Pauli{\mu} (\iu \Pauli{y}), \label{eq:Gam-proj}
\end{align}
where the Pauli matrices act in pseudospin space.
Since all three $t_{2g}$ orbitals are even, we may locally choose a gauge in which $V_{- \vb{k}} = V_{\vb{k}}$ so that $P_{\vb{k} a}^{\intercal} = s P_{-\vb{k} a} = s P_{\vb{k} a}$, where $\Gamma_a^{\intercal} = s \Gamma_a$.
$\mu = 0$ for antisymmetric $\Gamma_a$ ($s = -1$), whereas $\mu \in \{x, y, z\}$ for symmetric $\Gamma_a$ ($s = +1$).

Whenever a $g \in D_{4h}$ maps a $\vb{k}$ to itself modulo periodicity, its symmetry transformation matrix $U(g) = M(g) \otimes S(g)$ [Appendix~\ref{sec:SRO-TBA}, Table~\ref{tab:transf-mat}] commutes with the normal-state Hamiltonian $H_{\vb{k}}$:
\begin{align}
U^{\dag}(g) H_{\vb{k}} U(g) &= H_{R(g^{-1}) \vb{k}} = H_{\vb{k}}.
\end{align}
This means that the interband parts of $U(g)$ vanish.
As for the intraband part, we may always choose a basis for the Kramers degenerate subspace such that it takes a spin-like form:
\begin{align}
V_{\vb{k}}^{\dag} U(g) V_{\vb{k}} &= S(g).
\end{align}
The symmetry transformation rule of spin-orbit matrices [Eq.~\eqref{eq:transf-Gamma}, Appendix~\ref{sec:SC-construct}] now gives the constraint:
\begin{align}
S^{\dag}(g) P_{\vb{k} a} S^{*}(g) &= \sum_{b = 1}^{\dim \lambda} \rho_{ab}^{(\lambda)}(g) P_{\vb{k} b}.
\end{align}
For $\vb{k}$ on the Van Hove line $\left(0, \frac{\pi}{a}, k_3\right)$, the $g$ from Table~\ref{tab:rho-kvH} constrain certain $P_{a}^{\mu}(\vb{k})$ to vanish, depending on the (anti-)symmetry, irrep, and $k_3$.
The (anti-)symmetry $\Gamma_a^{\intercal} = s \Gamma_a$ we shall denote with an irrep superscript $s = \pm$.
Thus, for instance, $\Gamma \in A_{1g}^{-}$ are antisymmetric under transposition, whereas $\Gamma \in B_{1g}^{+}$ are symmetric under transposition.
The symmetry-enforced behavior of $P_{a}^{\mu}\!\left(0, \frac{\pi}{a}, k_3\right)$ we may summarize as follows:
\begin{itemize}
\item $\Gamma$ belonging to $A_{2g}^{-}$ and $B_{2g}^{-}$ have $P^{0} = 0$ for all $k_3$.
\item $\{\Gamma_1 \mid \Gamma_2\} \in E_g^{-}$ have $P_{2}^{0} = 0$ for all $k_3$, whereas $P_{1}^{0} = 0$ only at $k_3 = 0, \pm \frac{2 \pi}{c}$.
\item $\Gamma \in A_{1g}^{+}, B_{1g}^{+}$ have $P^{y} = P^{z} = 0$ for all $k_3$, and $P^{x} = 0$ only at $k_3 = 0, \pm \frac{2 \pi}{c}$.
\item $\Gamma \in A_{2g}^{+}, B_{2g}^{+}$ have $P^{x} = 0$ for all $k_3$, and $P^{y} = 0$ only at $k_3 = 0, \pm \frac{2 \pi}{c}$. $P^{z}$ is unconstrained.
\item $\{\Gamma_1 \mid \Gamma_2\} \in E_g^{+}$ have $P_{1}^{y} = P_{1}^{z} = P_{2}^{x} = 0$ for all $k_3$, and $P_{2}^{z} = 0$ only at $k_3 = 0, \pm \frac{2 \pi}{c}$. The remaining $P_{1}^{x}$ and $P_{2}^{y}$ are unconstrained.
\item The $P^{0}$ of $\Gamma$ from $A_{1g}^{-}$ and $B_{1g}^{-}$ are completely unconstrained for all $k_3$.
\end{itemize}
In the limit of vanishing body-centered tetragonal hopping, the following $P_{a}^{\mu}$ vanish in addition:
\begin{itemize}
\item For $\{\Gamma_1 \mid \Gamma_2\} \in E_g^{-}$, $P_{1}^{0}$ vanishes for all $k_3$ so both $P_{a}^{0}$ are zero.
\item For $\Gamma \in A_{1g}^{+}, B_{1g}^{+}$, $P^{\mu}$ completely vanish for all $k_3$.
\item For $\Gamma \in A_{2g}^{+}, B_{2g}^{+}$, $P^{y} = 0$ for all $k_3$, but $P^{z}$ is still unconstrained.
\item For $\{\Gamma_1 \mid \Gamma_2\} \in E_g^{+}$, $P_{2}^{z} = 0$ for all $k_3$, but $P_{1}^{x}$ and $P_{2}^{y}$ are still unconstrained.
\end{itemize}
Owning to the fact that all characteristically body-centered hopping is necessarily between layers and that these hoppings are very small in SRO because of its high anisotropy, the vanishing $P_{a}^{\mu}$ listed above are very small for SRO, although not precisely zero.
Using the tight-binding model of Ref.~\cite{Roising2019}, described in Appendix~\ref{sec:SRO-TBA}, we have quantified their smallness: the vanishing $P_{a}^{\mu}$ listed above are by a factor of $50$ or more smaller than the largest possible $P_{a}^{\mu} \sim 1$, where all $\Gamma_a$ have been normalized to $\tr \Gamma_a^{\dag} \Gamma_a = 1$ for a fair comparison.

Unlike the above anisotropy argument, arguments based on the $d_{xy}$ orbital character of the $\gamma$ band do not suppress any irreps, but only inform us on which $\Gamma_a$ from within a given irrep have large $P_{a}^{\mu}$.

Finally, we synthesize the results found for $d_a$ and $\Gamma_a$.
This is done by going through the multiplication table of irreps [Table~\ref{tab:irrep-mult} in Appendix~\ref{sec:SC-construct}] and seeing which entries yield a $\Delta_a(\vb{k})$ with a finite $\gamma$ band projection.
The results are summarized in Table~\ref{tab:main-result}.
Table~\ref{tab:main-result} is the main result of this paper.
As mentioned, SRO's anisotropy suppresses the blue entries of the table by two orders of magnitude.
This means that a $\Delta$ with a maximal value $\sim k_B T_c$ is way too small on the Van Hove lines to explain the observed entropy quenching~\cite{Li2022}.
Hence the blue entries of Table~\ref{tab:main-result} are excluded as possible leading SC states as well.

\begin{figure}[t]
\includegraphics[width=0.7\columnwidth]{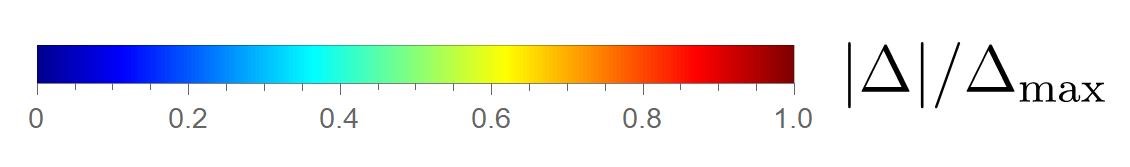}
\subfloat[$\Delta_{\vb{k}} = \Lambda_4 (\iu \Pauli{y}) \in A_{1g}$]{\includegraphics[width=\columnwidth]{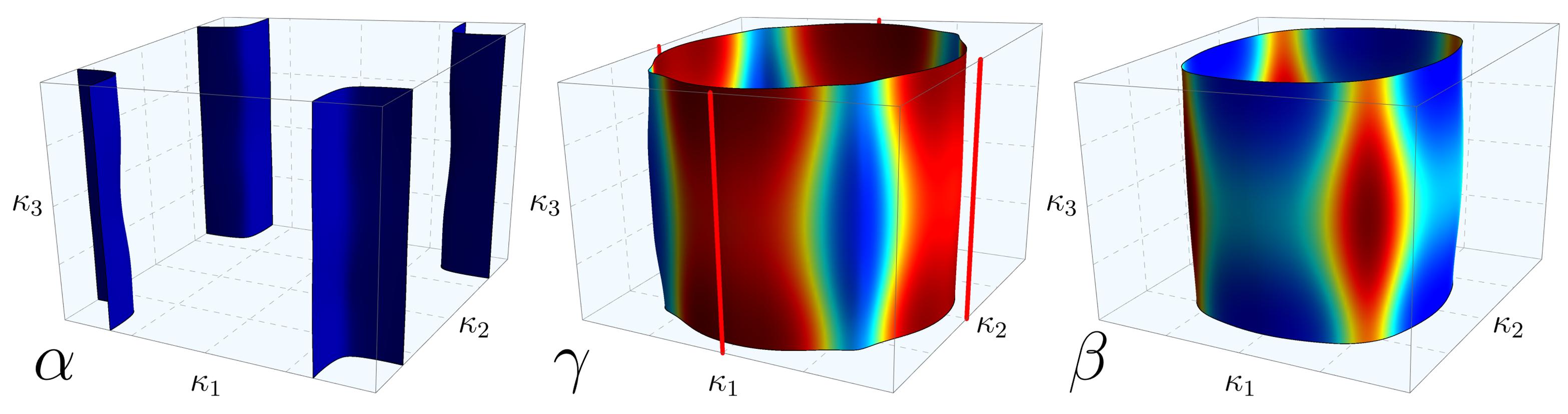}} \\
\subfloat[$\Delta_{\vb{k}} = \Lambda_4 (\iu \Pauli{y}) \left(\cos \kappa_1 - \cos \kappa_2\right) \in B_{1g}$]{\includegraphics[width=\columnwidth]{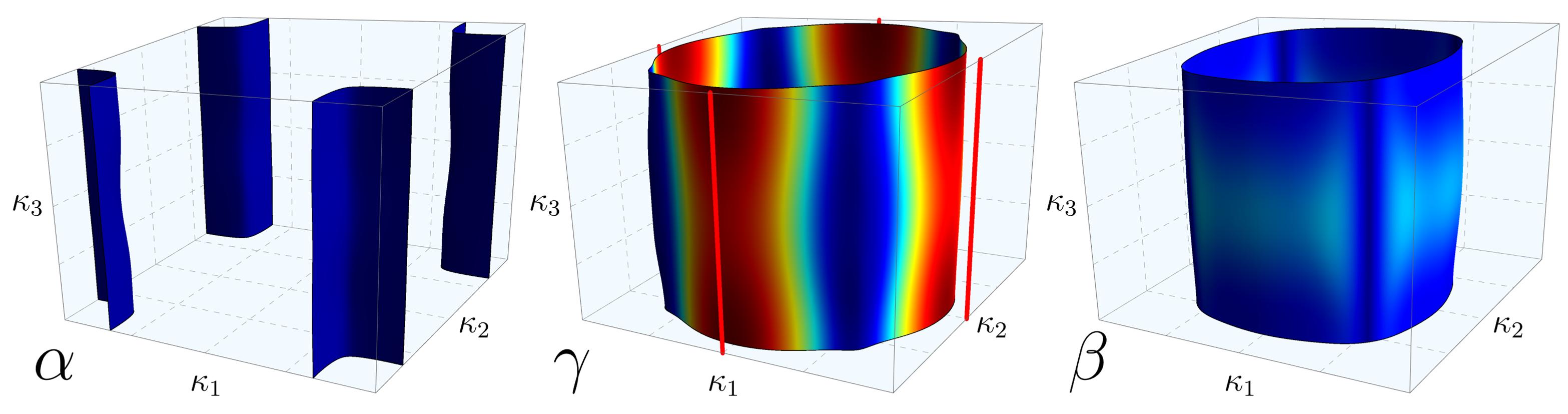}} \\
\subfloat[$\Delta_{\vb{k}} = \Lambda_4 (\iu \Pauli{y}) \left(\sin\frac{\kappa_1}{2} \cos\frac{\kappa_2}{2} \pm \iu \cos\frac{\kappa_1}{2} \sin\frac{\kappa_2}{2}\right) \sin\frac{\kappa_3}{2} \in E_g$]{\includegraphics[width=\columnwidth]{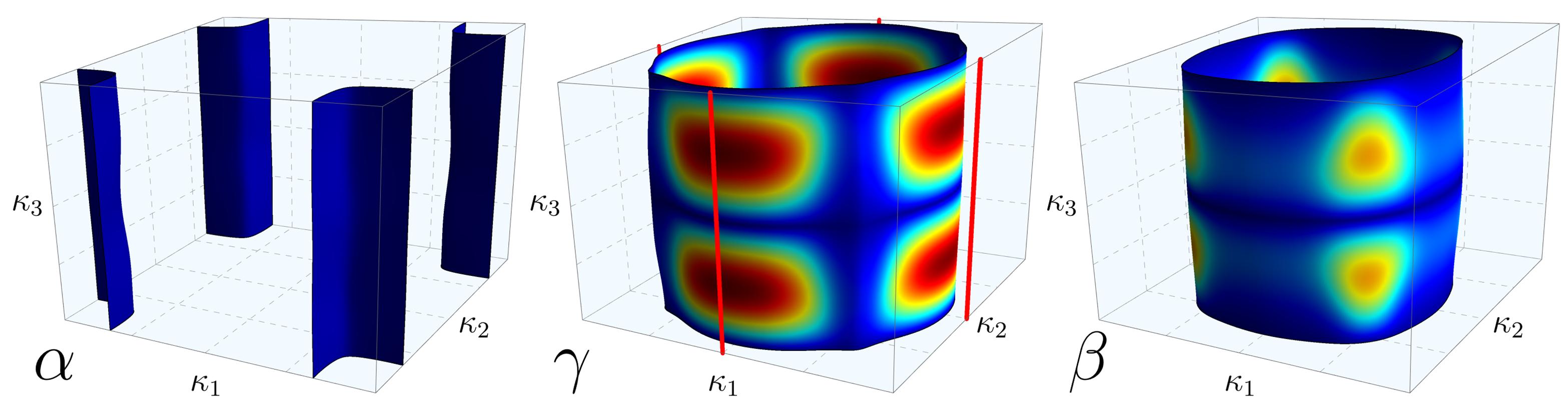}}
\caption{Projections onto the Fermi sheets of three Van Hove line-gapping SC states $\Delta_{\vb{k}}$ from Table~\ref{tab:main-result}, belonging to irreps $A_{1g}$~(a), $B_{1g}$~(b), and chiral $E_g$~(c), respectively.
$\kappa_1 = a k_1 \in [- \pi, \pi]$, $\kappa_2 = a k_2 \in [- \pi, \pi]$, and $\kappa_3 = c k_3 \in [- 2 \pi, 2 \pi]$.
In the $\gamma$ sheet plots, the Van Hove lines $\left(\pm \frac{\pi}{a}, 0, k_3\right)$ and $\left(0, \pm \frac{\pi}{a}, k_3\right)$ are highlighted red.
Because of the $\Lambda_4 = \diag(0, 0, \sqrt{2})$ orbital structure we have chosen, the projections are only large where the bands have $d_{xy}$ orbital character, which is on the $\gamma$ sheet and where the $\gamma$ and $\beta$ sheets nest.
More plots can be found in Appendix~\ref{sec:App-pics}.}
\label{fig:gaps-ex}
\end{figure}

From Table~\ref{tab:main-result} we see that, among even pairings, only $A_{1g}$, $B_{1g}$, and $E_g$ irreps have pairings that do not have symmetry-enforced vertical line nodes on the Van Hove lines.
Thus even pairings must have admixtures from one of these three irreps to be able to explain the elastocaloric experiment of Ref.~\cite{Li2022}.
It is worth noting that within these three irreps, pairings with symmetry-enforced vertical line nodes on $\vb{k}_{\text{VH}}$ do exist, like for instance $\Delta(\vb{k}) = \Lambda_1 (\iu \Pauli{y}) \sin a k_1 \sin a k_2 \in B_{2g}^{-} \otimes B_{2g} = A_{1g}$ [$\Lambda_1$ is given in Appendix~\ref{sec:GM-mat}].
So Table~\ref{tab:main-result} also yields non-trivial information on the spin-orbit and momentum structure of these Van Hove line-gapping admixtures.
Three representatives of such even-parity $\vb{k}_{\text{VH}}$-gapping SC states are plotted in Figure~\ref{fig:gaps-ex}.

\begin{table}[t]
\caption{Even-parity~(a) and odd-parity~(b) pairings that do not have a vertical line node at $\left(0, \frac{\pi}{a}, k_3\right)$, constructed by combining pairing wavefunctions $d_a(\vb{k})$ with spin-orbit matrices $\Gamma_a$ according to the multiplication table of $D_{4h}$ irreps [Table~\ref{tab:irrep-mult}, Appendix~\ref{sec:SC-construct}].
A `$+$' superscript on a spin-orbit matrix irrep means that the matrices are symmetric ($\Gamma^{\intercal} = + \Gamma$), whereas a `$-$' superscript indicates antisymmetry under transposition.
A zero component of $E_{g/u}$ means that it vanishes on $\left(0, \frac{\pi}{a}, k_3\right)$.
Highlighted red are those $d_a$ that must be periodic under body-centered translations, but not under simple tetragonal translations, to be finite on $\left(0, \frac{\pi}{a}, k_3\right)$.
For examples, see Table~\ref{tab:d-func-class} from Appendix~\ref{sec:SC-construct}.
Such $d_a$ have horizontal line nodes at $k_3 = 0, \pm \frac{2 \pi}{c}$.
Highlighted blue are those $\Gamma_a$ whose projections onto the $\gamma$ band are suppressed by two orders of magnitude because of the weakness of body-centered interlayer hopping.
Such $\Gamma_a$ are unable to account for the elastocaloric experiment, but are listed for the sake of completeness.}
\subfloat[Even pairings that are finite on $\left(0, \frac{\pi}{a}, k_3\right)$.]{
{\renewcommand{\arraystretch}{1.3}
\renewcommand{\tabcolsep}{5.4pt}
\begin{tabular}{|c"c|c|c|} \hline
$\otimes$ & $A_{1g}\{d\}$ & $B_{1g}\{d\}$ & $E_g\{\textcolor{red}{d_1} \mid 0\}$
\\ \thickhline
$A_{1g}^{-}\{\Gamma\}$ & $A_{1g}\{\Gamma d\}$ & $B_{1g}\{\Gamma d\}$ & $E_g\{\Gamma \textcolor{red}{d_1} \mid 0\}$
\\ \hline
$B_{1g}^{-}\{\Gamma\}$ & $B_{1g}\{\Gamma d\}$ & $A_{1g}\{\Gamma d\}$ & $E_g\{\Gamma \textcolor{red}{d_1} \mid 0\}$
\\ \hline
$E_g^{-}\begin{Bmatrix}\textcolor{blue}{\Gamma_1} \\ 0\end{Bmatrix}$
& $E_g\begin{Bmatrix}\textcolor{blue}{\Gamma_1} d \\ 0\end{Bmatrix}$
& $E_g\begin{Bmatrix}\textcolor{blue}{\Gamma_1} d \\ 0\end{Bmatrix}$
& $\begin{matrix}
A_{1g}\{\textcolor{blue}{\Gamma_1} \textcolor{red}{d_1} + 0\} \\
\oplus B_{1g}\{\textcolor{blue}{\Gamma_1} \textcolor{red}{d_1} - 0\}
\end{matrix}$
\\ \hline
\end{tabular}}} \\
\subfloat[Odd pairings that are finite on $\left(0, \frac{\pi}{a}, k_3\right)$.]{
{\renewcommand{\arraystretch}{1.3}
\renewcommand{\tabcolsep}{4.2pt}
\begin{tabular}{|c"c|c|c|} \hline
$\otimes$ & $A_{2u}\{d\}$ & $B_{2u}\{d\}$ & $E_u\{0 \mid \textcolor{red}{d_2}\}$
\\ \thickhline
$A_{1g}^{+}\{\textcolor{blue}{\Gamma}\}$ & $A_{2u}\{\textcolor{blue}{\Gamma} d\}$ & $B_{2u}\{\textcolor{blue}{\Gamma} d\}$ & $E_u\{0 \mid \textcolor{blue}{\Gamma} \textcolor{red}{d_2}\}$
\\ \hline
$A_{2g}^{+}\{\Gamma\}$ & $A_{1u}\{\Gamma d\}$ & $B_{1u}\{\Gamma d\}$ & $E_u\{\Gamma \textcolor{red}{d_2} \mid 0\}$
\\ \hline
$B_{1g}^{+}\{\textcolor{blue}{\Gamma}\}$ & $B_{2u}\{\textcolor{blue}{\Gamma} d\}$ & $A_{2u}\{\textcolor{blue}{\Gamma} d\}$ & $E_u\{0 \mid - \textcolor{blue}{\Gamma} \textcolor{red}{d_2}\}$
\\ \hline
$B_{2g}^{+}\{\Gamma\}$ & $B_{1u}\{\Gamma d\}$ & $A_{1u}\{\Gamma d\}$ & $E_u\{\Gamma \textcolor{red}{d_2} \mid 0\}$
\\ \hline
$E_g^{+}\begin{Bmatrix}\Gamma_1 \\ \Gamma_2\end{Bmatrix}$
& $E_u\begin{Bmatrix}\Gamma_2 d \\ - \Gamma_1 d\end{Bmatrix}$
& $E_u\begin{Bmatrix}\Gamma_2 d \\ \Gamma_1 d\end{Bmatrix}$
& $\begin{matrix}
A_{1u}\{0 + \Gamma_2 \textcolor{red}{d_2}\} \\
\oplus A_{2u}\{\Gamma_1 \textcolor{red}{d_2} - 0\} \\
\oplus B_{1u}\{0 - \Gamma_2 \textcolor{red}{d_2}\} \\
\oplus B_{2u}\{\Gamma_1 \textcolor{red}{d_2} + 0\}
\end{matrix}$
\\ \hline
\end{tabular}}}
\label{tab:main-result}
\end{table}

One such piece of information is that $E_g$ pairing must be made of wavefunctions $d_a$ that are body-centered periodic, but not simple tetragonal periodic.
The lowest order such $\{d_{xz} \mid d_{yz}\} \in E_g$ is:
\begin{equation}
\hspace{-2pt}
\left\{\sin\frac{a k_1}{2} \cos\frac{a k_2}{2} \sin\frac{c k_3}{2} \middle| \cos\frac{a k_1}{2} \sin\frac{a k_2}{2} \sin\frac{c k_3}{2}\right\}.
\label{eq:nastyone}
\end{equation}
It is this pairing state, only allowed because of the body-centered tetragonal structure of SRO, that opens a gap at the Van Hove line and that we cannot exclude based on the elastocaloric data.
In Ref.~\cite{Suh2020} it was shown that such a pairing state can be stabilized by a strongly momentum-dependent spin-orbit coupling.
A better understanding of the origin of such momentum dependence might help elucidate whether this state is a viable option for SRO's SC.
In distinction, the $E_g$ pairing state
\begin{equation}
\{\sin a k_1  \sin c k_3 \mid \sin a k_2  \sin c k_3 \}, \label{eq:usualone}
\end{equation}
which would be the only allowed one for simple-tetragonal lattices, cannot be the only pairing state as it does not open a gap on the Van Hove line.
An important difference between these two types of states [\eqref{eq:nastyone} vs.~\eqref{eq:usualone}] is that the former always have horizontal line nodes at $k_3 = 0, \pm \frac{2 \pi}{c}$.

Among odd pairings, all irreps have pairings without symmetry-enforced vertical line nodes on $\vb{k}_{\text{VH}}$.
However, the orientations of the Balian-Werthamer $\vb{d}$-vectors~\cite{Balian1963} are non-trivially restricted and the non-suppressed $A_{2u}$ and $B_{2u}$ pairings are necessarily made of characteristically body-centered periodic $d_a$.

In multiband systems with spin-orbit coupling, a $\vb{d}$-vector is associated with each band in its pseudospin (Kramers) space.
It is defined through:
\begin{align}
V_{\vb{k} n}^{\dag} \Delta(\vb{k}) V_{- \vb{k} n}^{*} &= \vb{d}_{\vb{k} n} \vdot \vb{\Pauli{}} (\iu \Pauli{y}),
\end{align}
where $V_{\vb{k} n} = \left(v_{\vb{k} n \uparrow}, v_{\vb{k} n \downarrow}\right)$ are the Kramers-degenerate eigenvectors of the $n$-th band and $V_{- \vb{k} n} = V_{\vb{k} n}$.
We make the following gauge choice for the pseudospins:
\begin{align}
\begin{aligned}
V_{\vb{k} n}^{\dag} (\one \otimes \iu \Pauli{y}) V_{\vb{k} n}^{*} &= \iu \Pauli{y}, \\
V_{\vb{k} n}^{\dag} (\one \otimes \Pauli{z}) V_{\vb{k} n} &= s_z \Pauli{z}, \\
V_{\vb{k} n}^{\dag} (\one \otimes \Pauli{x}) V_{\vb{k} n} &= s_x \Pauli{x} + \delta_{xz} \Pauli{z},
\end{aligned} \label{eq:pseudo-gauge}
\end{align}
where $s_z, s_x, \delta_{xz} \in \R$.
This is the closest one can make the pseudospins look like spins.
In general $\delta_{xz}$ is not zero, nor are the $\delta_{yx}, \delta_{yz}$ from $V_{\vb{k} n}^{\dag} (\one \otimes \Pauli{y}) V_{\vb{k} n} = s_y \Pauli{y} + \delta_{yx} \Pauli{x} + \delta_{yz} \Pauli{z}$.
However, in SRO the only regions where $\delta_{xz}, \delta_{yx}, \delta_{yz}$ are substantially different from zero is at the nesting of the $\alpha$, $\beta$, and $\gamma$ bands at $k_1 = \pm k_2$ [Figure~\ref{fig:bands}].
The explanation for this is the fact that spin-orbit coupling most strongly affects the band structure there.

Using the tight-binding model of SRO [Appendix~\ref{sec:SRO-TBA}], we have explored the orientation of the $\vb{d}_{\vb{k} n}$-vectors on the $\alpha$, $\beta$, and $\gamma$ Fermi sheets.
Everywhere except near the $k_1 = \pm k_2$ nesting of the sheets, we find that symmetric spin-orbit matrices from 1D irreps have $\vb{d}_{\vb{k} n}$ pointing along $\pm \vu{z}$, whereas $\{\Gamma_1 \mid \Gamma_2\}$ from $E_g^{+}$ always have in-plane $\vb{d}_{\vb{k} n}$.
So the non-suppressed $A_{2u}$ and $B_{2u}$ from Table~\ref{tab:main-result}~(b) have $\vb{d}_{\vb{k} n} \parallel \vu{z}$.
Moreover, among odd pairings not made of body-centered $d_a(\vb{k})$, $A_{1u}$ and $B_{1u}$ pairings have $\vb{d}_{\vb{k} n} \parallel \vu{z}$ and $E_u$ pairings have in-plane $\vb{d}_{\vb{k} n}$.
Given that body-centered $\{d_1 \mid d_2\} \in E_u$ have horizontal line nodes, on the one hand, and that the spin susceptibility is intimately related to the orientation of the Balian-Werthamer $\vb{d}$-vector, on the other, this information may prove to be useful in further narrowing down the odd-pairing SC candidates.

\section{Conclusion}
This paper was motivated by the measurements of the elastocaloric effect of \ce{Sr2RuO4} under strain reported in Ref.~\cite{Li2022}.
The elastocaloric effect measures, with high accuracy, the entropy derivative $\partial S(\varepsilon, T) / \partial \varepsilon$.
Above $T_c$, the elastocaloric effect revealed a pronounced maximum in the entropy as function of strain $\varepsilon$.
As demonstrated in Ref.~\cite{Li2022}, this maximum of $S(\varepsilon)$ can be fully accounted for by the DOS enhancement that occurs when the Fermi energy crosses the Van Hove points near the lines $\left(0, \pm \frac{\pi}{a}, k_3\right)$.
Below $T_c$, the entropy maximum was found to transform into a minimum.
This is only possible if the states near the saddle points of the electronic dispersion open a gap as one enters the SC state.
Hence, with rather minimal modelling, it is possible to obtain information about the momentum-space structure of the SC gap from a thermodynamic measurement.

In order to draw more detailed conclusions about the allowed pairing states, we performed a symmetry analysis for a three-dimensional, three-band description of SRO.
Here we focus primarily on even-parity states, given the strong evidence for even parity in NMR measurements~\cite{Pustogow2019, Ishida2020, Chronister2021}.
From a simple two-dimensional perspective, one would conclude that the SC state must open a gap at the Van Hove points $\left(\pm \frac{\pi}{a}, 0\right)$ and $\left(0, \pm\frac{\pi}{a}\right)$.
However, to distinguish the relevant pairing states, in particular those of the 2D irreducible representation $E_g$ that transform like $\{d_{xz} \mid d_{yz}\}$, we must include the third momentum direction.
It is well known that the energy dispersion of SRO is strongly anisotropic.
Indeed, our analysis shows that the energy scale below which the three-dimensionality of the Fermi surface becomes important is about one kelvin, fully consistent with magneto-oscillation experiments~\cite{Mackenzie2003}.
We also show that the saddle points deviate by very small amounts $\var{k_{\text{VH}, 2}} \ll \frac{2\pi}{a}$ from the lines $\left(\pm \frac{\pi}{a}, 0, k_3\right)$ and $\left(0, \pm \frac{\pi}{a}, k_3\right)$.
However, this need not be the case for the SC state.
While the single particle spectrum of SRO is highly anisotropic, it is possible that many-body interactions that are responsible for the SC pairing couple different layers more efficiently.
Hence, at least in principle, one should not exclude a strong dependence of the gap function on $k_3$; such dependence is crucial for the $\{d_{xz} \mid d_{yz}\}$-wave pairing states.

With these insights, we then turned to the symmetry analysis of potential pairing states.
If one assumes for a moment that the crystal structure of SRO is simple tetragonal, one is left with only two possible even pairing states, namely, the $s$-wave state of $A_{1g}$ symmetry and the $d_{x^2-y^2}$-wave state of $B_{1g}$ symmetry.
Given that fine-tuning is required for $s$-wave pairing to be consistent with the pair-breaking role of impurities~\cite{Mackenzie1998, *Mackenzie1998-E, Mao1999, Kikugawa2002, Kikugawa2004}, $d_{x^2-y^2}$-wave pairing would then appear to be the only natural pairing candidate.
However, \ce{Sr2RuO4} is a body-centered tetragonal compound.
The corresponding symmetry analysis now allows, in addition to $d_{x^2-y^2}$-wave pairing, for a $\{d_{xz} \mid d_{yz}\}$-wave state of $E_g$ symmetry like the one given in Eq.~\eqref{eq:nastyone}.

Our analysis does, however, allow us to exclude $d_{xy}$-wave pairing states that transform like $B_{2g}$ and $g_{xy(x^2-y^2)}$-wave pairing states that transform like $A_{2g}$ as sole pairing states.
Such states may at best be subleading contenders that could be added to the pairing wavefunction at fine-tuned points of accidental degeneracy.
In addition, we can exclude $\{d_{xz} \mid d_{yz}\}$-wave pairing that is exclusively of the type given in Eq.~\eqref{eq:usualone}.
The nature of our argument does not allow us to more precisely quantify how large these subleading $d_{xy}$-wave or $g_{xy(x^2-y^2)}$-wave contributions are because they vanish precisely where the elastocaloric experiment is most sensitive: at the Van Hove lines.
Thus, while the elastocaloric measurements do not allow for a unique determination of the superconducting order parameter symmetry, they do constrain the available options.
To finally resolve the nature of superconductivity in \ce{Sr2RuO4} requires a better understanding of the origin of time-reversal symmetry breaking and of the orientation of line nodes.

\begin{acknowledgments}
We are grateful to Markus Garst for helpful discussions.
This work was supported by the Deutsche Forschungsgemeinschaft (DFG, German Research Foundation) -- TRR 288-422213477 Elasto-Q-Mat project A05 (R.V.), project A07 (G.P.\ and J.S.), and project A10 (C.H.\ and A.P.M.).
A.R.\ acknowledges support from the Engineering and Physical Sciences Research Council (grant numbers EP/P024564/1, EP/S005005/1, and EP/V049410/1).
Research in Dresden benefits from the environment provided by the DFG Cluster of Excellence ct.qmat (EXC 2147, project ID 390858940).
\end{acknowledgments}

\appendix

\section{Gell-Mann matrices} \label{sec:GM-mat}
We use the following unconventional choice for the nine Gell-Mann matrices:
\begin{align}
\Lambda_0 &= \begin{pmatrix}
1 & 0 & 0 \\
0 & 1 & 0 \\
0 & 0 & 0
\end{pmatrix}, & \Lambda_1 &= \begin{pmatrix}
0 & 1 & 0 \\
1 & 0 & 0 \\
0 & 0 & 0
\end{pmatrix}, \\
\Lambda_2 &= \begin{pmatrix}
0 & -\iu & 0 \\
\iu & 0 & 0 \\
0 & 0 & 0
\end{pmatrix}, & \Lambda_3 &= \begin{pmatrix}
1 & 0 & 0 \\
0 & -1 & 0 \\
0 & 0 & 0
\end{pmatrix}, \\
\Lambda_4 &= \begin{pmatrix}
0 & 0 & 0 \\
0 & 0 & 0 \\
0 & 0 & \sqrt{2}
\end{pmatrix},
\end{align}
and:
\begin{align}
\Lambda_5 &= \begin{pmatrix}
0 & 0 & 1 \\
0 & 0 & 0 \\
1 & 0 & 0
\end{pmatrix}, & \Lambda_6 &= \begin{pmatrix}
0 & 0 & -\iu \\
0 & 0 & 0 \\
\iu & 0 & 0
\end{pmatrix}, \\
\Lambda_7 &= \begin{pmatrix}
0 & 0 & 0 \\
0 & 0 & 1 \\
0 & 1 & 0
\end{pmatrix}, & \Lambda_8 &= \begin{pmatrix}
0 & 0 & 0 \\
0 & 0 & -\iu \\
0 & \iu & 0
\end{pmatrix}.
\end{align}
They are normalized so that $\tr \Lambda_{\mu} \Lambda_{\nu} = 2 \Kd_{\mu \nu}$.

\section{Tight-binding model of SRO} \label{sec:SRO-TBA}
Within the effective tight-binding model of SRO based on the $t_{2g}$ orbitals of \ce{Ru}, the point group operation $g \in D_{4h}$ acts on electrons according to:
\begin{align}
\hat{\mathcal{U}}^{\dag}(g) \psi_{\vb{k}} \hat{\mathcal{U}}(g) &= M(g) \otimes S(g) \psi_{R(g^{-1}) \vb{k}},
\end{align}
where $\psi_{\vb{k}}$ are column vectors of fermionic destruction operators in the basis $\left(d_{yz \uparrow}, d_{yz \downarrow}, d_{zx \uparrow}, d_{zx \downarrow}, d_{xy \uparrow}, d_{xy \downarrow}\right)^{\intercal}$, $\hat{\mathcal{U}}(g)$ are the Fock-space symmetry operators, and $R, M, S$ are unitary representations of $D_{4h}$ whose generators are listed in Table~\ref{tab:transf-mat}.
Time-reversal $\Theta$ acts like:
\begin{align}
\Theta^{-1} \psi_{\vb{k}} \Theta &= (\one \otimes \iu \Pauli{y}) \psi_{- \vb{k}},
\end{align}
where $\one$ is the $3 \times 3$ identity matrix and $\Pauli{\mu}$ are the Pauli matrices.

Since there is only one ruthenium atom per a body-centered unit cell, the tight-binding Hamiltonian takes the form:
\begin{equation}
H_0 = - \sum_{\vb{R}, \vb{\delta}} \psi_{\vb{R} + \vb{\delta}}^{\dag} \left[T_{\vb{\delta}} \otimes \one + \iu \sum_{\mu=1}^{3} \Lambda_{\vb{\delta}; \mu} \otimes \Pauli{\mu}\right] \psi_{\vb{R}},
\end{equation}
where $\vb{R}, \vb{\delta}$ go over the body-centered tetragonal lattice whose primitive lattice vectors are:
\begin{equation}
\vb{a}_1 = a \vu{e}_1, \;\;
\vb{a}_2 = a \vu{e}_2, \;\;
\vb{a}_3 = \tfrac{1}{2}\!\left(a \vu{e}_1 + a \vu{e}_2 + c \vu{e}_3\right).
\end{equation}
The Hamiltonian is hermitian only when $T_{- \vb{\delta}} = T_{\vb{\delta}}^{\dag}$ and $\Lambda_{- \vb{\delta}; \mu} = - \Lambda_{\vb{\delta}; \mu}^{\dag}$.
Point group symmetries constrain and relate different hopping amplitudes:
\begin{gather}
M^{\dag}(g) T_{\vb{\delta}} M(g) = T_{R(g^{-1}) \vb{\delta}}, \\
M^{\dag}(g) \Lambda_{\vb{\delta}; \mu} M(g) = \det R(g) \sum_{\nu=1}^{3} R_{\mu\nu}(g) \Lambda_{R(g^{-1}) \vb{\delta}; \nu}.
\end{gather}
To ensure time-reversal invariance, all matrix elements must be made real, i.e., $T_{\vb{\delta}}^{*} = T_{\vb{\delta}}$ and $\Lambda_{\vb{\delta}; \mu}^{*} = \Lambda_{\vb{\delta}; \mu}$.

\begin{table}[t]
\caption{The generators of the representations $R$, $M$, and $S$ of the point group $D_{4h}$.
$C_{4z}$ is a rotation by $\pi/2$ around $z$.
$C_{2x}'$ and $C_{2d}''$ are rotations by $\pi$ around $x$ and the diagonal $x+y$, respectively.
$P$ is parity.}
{\renewcommand{\arraystretch}{1.3}
\renewcommand{\tabcolsep}{8.4pt}
\begin{tabular}{cccc} \hline\hline
$g$ & $R(g)$ & $M(g)$ & $S(g)$ 
\\ \hline
$C_{4z}$ & $\begin{pmatrix} 0 & -1 & 0 \\ 1 & 0 & 0 \\ 0 & 0 & 1 \end{pmatrix}$ & $\begin{pmatrix} 0 & 1 & 0 \\ -1 & 0 & 0 \\ 0 & 0 & -1 \end{pmatrix}$ & $\displaystyle \frac{\Pauli{0} - \iu \Pauli{z}}{\sqrt{2}}$ \\
$C_{2x}'$ & $\begin{pmatrix} 1 & 0 & 0 \\ 0 & -1 & 0 \\ 0 & 0 & -1 \end{pmatrix}$ & $\begin{pmatrix} 1 & 0 & 0 \\ 0 & -1 & 0 \\ 0 & 0 & -1 \end{pmatrix}$ & $\displaystyle - \iu \Pauli{x}$ \\
$C_{2d}''$ & $\begin{pmatrix} 0 & 1 & 0 \\ 1 & 0 & 0 \\ 0 & 0 & -1 \end{pmatrix}$ & $\begin{pmatrix} 0 & -1 & 0 \\ -1 & 0 & 0 \\ 0 & 0 & 1 \end{pmatrix}$ & $\displaystyle - \iu \frac{\Pauli{x} + \Pauli{y}}{\sqrt{2}}$ \\
$P$ & $- \one$ & $\one$ & $\displaystyle \Pauli{0}$
\\ \hline\hline
\end{tabular}}
\label{tab:transf-mat}
\end{table}

\begin{table}[t]
\caption{Upper part: the values of our tight-binding model parameters in \si{\milli\electronvolt} according to Ref.~\cite{Roising2019}.
The parameters set to zero have not been considered in Ref.~\cite{Roising2019}.
Lower part: the parameters in \si{\milli\electronvolt} according to other references.
Those parameters not listed vanish.}
{\renewcommand{\arraystretch}{1.3}
\renewcommand{\tabcolsep}{3.3pt}
\begin{tabular}{c|c|c|c|c|c|c|c|c} \hline\hline
$t_1$ & $t_2$ & $t_3$ & $t_4$ & $t_5$ & $t_6$ & $t_7$ & $t_8$ & $t_9$ \\
$27.8$ & $257.8$ & $-22.4$ & $13.6$ & $3.2$ & $-35.5$ & $0$ & $-4.7$ & $0$
\\ \hline
$t_{10}$ & $t_{11}$ & $\bar{t}_1$ & $\bar{t}_2$ & $\bar{t}_3$ & $\bar{t}_4$ & $\bar{t}_5$ & $\bar{t}_6$ & $\bar{t}_7$ \\
$0$ & $-2.4$ & $356.8$ & $126.3$ & $-1.0$ & $17.0$ & $22.3$ & $0$ & $0$
\\ \hline
$\mu_{\text{1D}}$ & $\mu_{\text{2D}}$ & $t_{i1}$ & $t_{i2}$ & $t_{i3}$ & $t_{i4}$ & $t_{j}$ & $\eta_1$ & $\eta_2$ \\
$286.9$ & $351.9$ & $-2.0$ & $7.8$ & $0$ & $0$ & $2.7$ & $59.2$ & $59.2$
\\ \hline\hline
\end{tabular}}
{\renewcommand{\arraystretch}{1.3}
\renewcommand{\tabcolsep}{5.0pt}
\begin{tabular}{c|c|c|c|c|c|c|c|c|c|c}
$t_1$ & $t_2$ & $\bar{t}_1$ & $\bar{t}_2$ & $\bar{t}_4$ & $\mu_{\text{1D}}$ & $\mu_{\text{2D}}$ & $t_{i1}$ & $\eta_1$ & $\eta_2$ & Ref.
\\ \hline
$16$ & $145$ & $81$ & $39$ & $5$ & $122$ & $122$ & $0$ & $32$ & $32$ & \cite{Zabolotnyy2013}
\\ \hline
$9$ & $88$ & $80$ & $40$ & $5$ & $109$ & $109$ & $0$ & $35$ & $35$ & \cite{Cobo2016}
\\ \hline
$13$ & $165$ & $119$ & $49$ & $0$ & $178$ & $176$ & $21$ & $0$ & $0$ & \cite{Burganov2016}
\\ \hline\hline
\end{tabular}}
\label{tab:TBA-params}
\end{table}

Symmetries that map $\vb{\delta}$ to itself constrain the forms of the hopping amplitudes.
For the eight closest $\vb{\delta}$ of SRO, we find that:
\begin{align}
&T_{\vb{0}} = \begin{pmatrix} \mu_{\text{1D}} & 0 & 0 \\ 0 & \mu_{\text{1D}} & 0 \\ 0 & 0 & \mu_{\text{2D}} \end{pmatrix}, \;\;
T_{\vb{a}_1} = \begin{pmatrix} t_1 & 0 & 0 \\ 0 & t_2 & 0 \\ 0 & 0 & \bar{t}_1 \end{pmatrix}, \\
&T_{\vb{a}_1 + \vb{a}_2} = \begin{pmatrix} t_3 & t_{i1} & 0 \\ t_{i1} & t_3 & 0 \\ 0 & 0 & \bar{t}_2 \end{pmatrix}, \;\;
T_{\vb{a}_3} = \begin{pmatrix} t_4 & t_{i2} & t_{j} \\ t_{i2} & t_4 & t_{j} \\ t_{j} & t_{j} & \bar{t}_3 \end{pmatrix}, \\
&T_{2 \vb{a}_1} = \begin{pmatrix} t_5 & 0 & 0 \\ 0 & t_6 & 0 \\ 0 & 0 & \bar{t}_4 \end{pmatrix}, \;\;
T_{2 \vb{a}_1 + \vb{a}_2} = \begin{pmatrix} t_7 & t_{i3} & 0 \\ t_{i3} & t_8 & 0 \\ 0 & 0 & \bar{t}_5 \end{pmatrix}, \\
&T_{2 (\vb{a}_1 + \vb{a}_2)} = \begin{pmatrix} t_9 & t_{i4} & 0 \\ t_{i4} & t_9 & 0 \\ 0 & 0 & \bar{t}_6 \end{pmatrix}, \;\;
T_{3 \vb{a}_1} = \begin{pmatrix} t_{10} & 0 & 0 \\ 0 & t_{11} & 0 \\ 0 & 0 & \bar{t}_7 \end{pmatrix}.
\end{align}
Among these closest and thus largest $T_{\vb{\delta}}$, only $T_{\vb{a}_3}$ connects different layers, reflecting the high anisotropy of SRO.
Moreover, it is only through $T_{\vb{a}_3}$ that the body-centered periodicity of SRO is felt on the level of the one-particle Hamiltonian.
The on-site spin-orbit coupling takes the form:
\begin{gather}
\Lambda_{\vb{0}; 1} = \begin{pmatrix} 0 & 0 & 0 \\ 0 & 0 & - \eta_1 \\ 0 & \eta_1 & 0 \end{pmatrix}, \;\;
\Lambda_{\vb{0}; 2} = \begin{pmatrix} 0 & 0 & \eta_1 \\ 0 & 0 & 0 \\ - \eta_1 & 0 & 0 \end{pmatrix}, \\
\Lambda_{\vb{0}; 3} = \begin{pmatrix} 0 & - \eta_2 & 0 \\ \eta_2 & 0 & 0 \\ 0 & 0 & 0 \end{pmatrix}.
\end{gather}
Off-site ($\vb{k}$-dependent) spin-orbit coupling we shall not include, although one should keep in mind that some~\cite{Suh2020} have found that it has a large effect on the preferred Cooper pairing, even when small.

For our analysis, we have used the tight-binding parameter values of Ref.~\cite{Roising2019}, which they found by fitting to the ARPES-based tight-binding $17$-band model of Ref.~\cite{Veenstra2014}.
Their tight-binding parameter values are reproduced in Table~\ref{tab:TBA-params}.
The hopping amplitudes of Refs.~\cite{Roising2019} and~\cite{Suh2020} are broadly in agreement, as one would expect given that both were fitted to Ref.~\cite{Veenstra2014}.
However, the hoppings of both~\cite{Roising2019, Suh2020} are by a factor of two or so larger than those of Refs.~\cite{Zabolotnyy2013, Cobo2016, Burganov2016}, which are also ARPES-derived; see Table~\ref{tab:TBA-params}.
Although all these models give the correct shapes for the Fermi sheets, find that the $\gamma$ band is responsible for over \SI{50}{\percent} of the normal-state DOS, and predict a roughly \SI{20}{\percent} increase in the DOS at Van Hove strain, consistent with our entropy data [Figure~\ref{fig:elasto}], the predicted values for the total DOS differ by a factor of two.
Only Ref.~\cite{Zabolotnyy2013} has checked that their model gives a total DOS ($g \approx 17$ states per \si{\electronvolt} per body-centered tetragonal unit cell) that is consistent with the experimentally measured Sommerfeld coefficient $\gamma = (\pi^2 / 3) R g \approx \SI[per-mode=symbol]{40}{\milli\joule\per\square\kelvin\per\mol}$~\cite{NishiZaki2000, Deguchi2004, Kittaka2018}, where $R$ is the molar gas constant.
The main takeaway is that the various estimates cited in the main text might be off by a factor of two, which is still sufficient for our purposes and does not impact our argument in any way.

\begin{widetext}
In momentum space, the tight-binding Hamiltonian equals:
\begin{align}
H_{\vb{k}} &= - \sum_{\vb{\delta}} \left[T_{\vb{\delta}} \otimes \one + \iu \sum_{\mu=1}^{3} \Lambda_{\vb{\delta}; \mu} \otimes \Pauli{\mu}\right] \Elr^{- \iu \vb{k} \vdot \vb{\delta}} \\
&= \begin{pmatrix}
\epsilon_{\text{1D}}(\vb{k}) & \epsilon_{i}(\vb{k}) & \epsilon_{j}(\vb{k}) \\
\epsilon_{i}(\vb{k}) & \tilde{\epsilon}_{\text{1D}}(\vb{k}) & \tilde{\epsilon}_{j}(\vb{k}) \\
\epsilon_{j}(\vb{k}) & \tilde{\epsilon}_{j}(\vb{k}) & \epsilon_{\text{2D}}(\vb{k})
\end{pmatrix} \otimes \Pauli{0} + \begin{pmatrix}
0 & \iu \eta_2 \Pauli{3} & - \iu \eta_1 \Pauli{2} \\
- \iu \eta_2 \Pauli{3} & 0 & \iu \eta_1 \Pauli{1} \\
\iu \eta_1 \Pauli{2} & - \iu \eta_1 \Pauli{1} & 0
\end{pmatrix},
\end{align}
where $\tilde{\epsilon}_{\text{1D}}(k_1, k_2, k_3) = \epsilon_{\text{1D}}(k_2, k_1, k_3)$, $\tilde{\epsilon}_{j}(k_1, k_2, k_3) = \epsilon_{j}(k_2, k_1, k_3)$, and:
\begin{align}
\epsilon_{\text{1D}}(\vb{k}) &= - \mu_{\text{1D}} - 2 t_1 \cos \kappa_1 - 2 t_2 \cos \kappa_2 - 4 t_3 \cos \kappa_1 \cos \kappa_2 - 8 t_4 \cos \tfrac{1}{2} \kappa_1 \cos \tfrac{1}{2} \kappa_2 \cos \tfrac{1}{2} \kappa_3 - 2 t_5 \cos 2 \kappa_1 - 2 t_6 \cos 2 \kappa_2 \notag \\
&\qquad - 4 t_7 \cos 2 \kappa_1 \cos \kappa_2 - 4 t_8 \cos \kappa_1 \cos 2 \kappa_2 - 4 t_9 \cos 2 \kappa_1 \cos 2 \kappa_2 - 2 t_{10} \cos 3 \kappa_1 - 2 t_{11} \cos 3 \kappa_2, \\
\epsilon_{\text{2D}}(\vb{k}) &= - \mu_{\text{2D}} - 2 \bar{t}_1 \left(\cos \kappa_1 + \cos \kappa_2\right) - 4 \bar{t}_2 \cos \kappa_1 \cos \kappa_2 - 8 \bar{t}_3 \cos \tfrac{1}{2} \kappa_1 \cos \tfrac{1}{2} \kappa_2 \cos \tfrac{1}{2} \kappa_3 - 2 \bar{t}_4 \left(\cos 2 \kappa_1 + \cos 2 \kappa_2\right) \notag \\
&\qquad - 4 \bar{t}_5 \left(\cos 2 \kappa_1 \cos \kappa_2 + \cos \kappa_1 \cos 2 \kappa_2\right) - 4 \bar{t}_6 \cos 2 \kappa_1 \cos 2 \kappa_2 - 2 \bar{t}_7 \left(\cos 3 \kappa_1 + \cos 3 \kappa_2\right), \\
\epsilon_{i}(\vb{k}) &= 4 t_{i1} \sin \kappa_1 \sin \kappa_2 + 8 t_{i2} \sin \tfrac{1}{2} \kappa_1 \sin \tfrac{1}{2} \kappa_2 \cos \tfrac{1}{2} \kappa_3 + 8 t_{i3} \left(\cos \kappa_1 + \cos \kappa_2\right) \sin \kappa_1 \sin \kappa_2 + 4 t_{i4} \sin 2 \kappa_1 \sin 2 \kappa_2, \\
\epsilon_{j}(\vb{k}) &= 8 t_{j} \sin \tfrac{1}{2} \kappa_1 \cos \tfrac{1}{2} \kappa_2 \sin \tfrac{1}{2} \kappa_3.
\end{align}
Above $\vb{k} = (k_1, k_2, k_3)$, $\kappa_1 = a k_1$, $\kappa_2 = a k_2$, and $\kappa_3 = c k_3$.
\end{widetext}

The coupling to strain, needed for Figure~\ref{fig:bands}, was taken from the Supplementary information of Ref.~\cite{Li2022}.
The dispersion of the $\gamma$ band near the Van Hove line $\left(0, \frac{\pi}{a}, k_3\right)$, provided in the main text in Eqs.~\eqref{eq:saddle-epsilon} and~\eqref{eq:VHdispexpansion}, was found by diagonalizing $H_{\vb{k}}$ with the parameter values of Ref.~\cite{Roising2019}.

\begin{table}[t!]
\caption{A sample of possible pairing wavefunctions $d_a(\vb{k})$ categorized according to the transformation rule~\eqref{eq:transf-d}.
The irrep subscripts $g$ and $u$ mean even and odd under parity, respectively.
The two-component $\{d_1(\vb{k}) \mid d_2(\vb{k})\}$ transform according to the $\rho^{(E)}(g)$ given in Eq.~\eqref{eq:E-rep-rho}.
$\vb{k} = (k_1, k_2, k_3)$ and $\kappa_1 = a k_1$, $\kappa_2 = a k_2$, $\kappa_3 = c k_3$.
Highlighted red are those wavefunctions that are periodic under body-centered translations, but not under simple tetragonal translations.}
{\renewcommand{\arraystretch}{1.3}
\renewcommand{\tabcolsep}{5.4pt}
\begin{tabular}{c|c} \hline\hline
$A_{1g}$ & $1$, $\cos \kappa_1 + \cos \kappa_2$, $\cos \kappa_3$, $\cos \kappa_1 \cos \kappa_2$
\\ \hline
$A_{2g}$ & $\left(\cos \kappa_1 - \cos \kappa_2\right) \sin \kappa_1 \sin \kappa_2$
\\ \hline
$B_{1g}$ & $\cos \kappa_1 - \cos \kappa_2$
\\ \hline
$B_{2g}$ & $\sin \kappa_1 \sin \kappa_2$, $\textcolor{red}{\sin \tfrac{1}{2} \kappa_1 \sin \tfrac{1}{2} \kappa_2 \cos \tfrac{1}{2} \kappa_3}$
\\ \hline
$E_g$ & $\begin{matrix}
\left\{\sin \kappa_2 \sin \kappa_3 \mid - \sin \kappa_1 \sin \kappa_3\right\}, \\
\textcolor{red}{\left\{\cos \tfrac{1}{2} \kappa_1 \sin \tfrac{1}{2} \kappa_2 \sin\tfrac{1}{2} \kappa_3 \mid - \sin\tfrac{1}{2} \kappa_1 \cos\tfrac{1}{2} \kappa_2 \sin\tfrac{1}{2} \kappa_3\right\}}
\end{matrix}$
\\ \hline\hline
$A_{1u}$ & $\textcolor{red}{\left(\cos \kappa_1 - \cos \kappa_2\right) \sin \tfrac{1}{2} \kappa_1 \sin \tfrac{1}{2} \kappa_2 \sin \tfrac{1}{2} \kappa_3}$
\\ \hline
$A_{2u}$ & $\sin \kappa_3$, $\textcolor{red}{\cos \tfrac{1}{2} \kappa_1 \cos \tfrac{1}{2} \kappa_2 \sin \tfrac{1}{2} \kappa_3}$
\\ \hline
$B_{1u}$ & $\textcolor{red}{\sin \tfrac{1}{2} \kappa_1 \sin \tfrac{1}{2} \kappa_2 \sin \tfrac{1}{2} \kappa_3}$
\\ \hline
$B_{2u}$ & $\left(\cos \kappa_1 - \cos \kappa_2\right) \sin \kappa_3$
\\ \hline
$E_u$ & $\begin{matrix}
\left\{\sin \kappa_1 \mid \sin \kappa_2\right\}, \\
\left\{\left(\cos \kappa_1 - \cos \kappa_2\right) \sin \kappa_1 \mid - \left(\cos \kappa_1 - \cos \kappa_2\right) \sin \kappa_2\right\}, \\
\textcolor{red}{\left\{\sin \tfrac{1}{2} \kappa_1 \cos \tfrac{1}{2} \kappa_2 \cos \tfrac{1}{2} \kappa_3 \mid \cos \tfrac{1}{2} \kappa_1 \sin \tfrac{1}{2} \kappa_2 \cos \tfrac{1}{2} \kappa_3\right\}}
\end{matrix}$
\\ \hline\hline
\end{tabular}}
\label{tab:d-func-class}
\end{table}

\section{Superconducting states of SRO} \label{sec:SC-construct}
For the purpose of classifying even-frequency pairings, it is sufficient to consider the static case because the two behave the same symmetry-wise~\cite{LinderBalatsky2019}.
Odd-frequency pairings are beyond the scope of this article.
On the mean-field level, static zero-momentum SC is described by a pairing term in the Hamiltonian of the form:
\begin{align}
H_{\text{SC}} &= \sum_{\vb{k} \alpha \beta} \psi_{\vb{k} \alpha}^{\dag} \Delta_{\alpha \beta}(\vb{k}) \psi_{- \vb{k} \beta}^{\dag} + \hc,
\end{align}
where $\alpha, \beta$ are spin-orbit indices.
Because of the fermionic anticommutation, the SC gap matrix $\Delta_{\alpha \beta}(\vb{k})$ satisfies the exchange property:
\begin{align}
\Delta^{\intercal}(\vb{k}) &= - \Delta(- \vb{k}), \label{eq:exch-prop}
\end{align}
where ${}^{\intercal}$ is transposition.

If the pairing were conventional, all point group operations would be preserved and $\hat{\mathcal{U}}^{\dag}(g) H_{\text{SC}} \hat{\mathcal{U}}(g) = H_{\text{SC}}$ would hold for all $g \in D_{4h}$, giving the constraint $U^{\dag}(g) \Delta\!\left(R(g) \vb{k}\right)  U^{*}(g) = \Delta(\vb{k})$, where $U(g) \equiv M(g) \otimes S(g)$.
Unconventional pairing is classified by the way it breaks this constraint:
\begin{align}
U^{\dag}(g) \Delta_a\!\left(R(g) \vb{k}\right)  U^{*}(g) &= \sum_{b = 1}^{\dim \lambda} \rho_{ab}^{(\lambda)}(g) \Delta_b(\vb{k}). \label{eq:transf-Delta}
\end{align}
Here, $\lambda$ is an irrep of $D_{4h}$, $a, b$ are indices internal to the irrep, and $\rho_{ab}^{(\lambda)}$ are the corresponding matrices.
Only for the 2D irreps $E_{g/u}$ are there multiple possible $\rho_{ab}^{(\lambda)}$.
We choose (cf.~representation $R$):
\begin{align}
\begin{aligned}
\rho^{(E)}(C_{4z}) &= \begin{pmatrix} 0 & -1 \\ 1 & 0 \end{pmatrix}, &
\rho^{(E)}(C_{2x}') &= \begin{pmatrix} 1 & 0 \\ 0 & -1 \end{pmatrix}, \\
\rho^{(E)}(C_{2d}'') &= \begin{pmatrix} 0 & 1 \\ 1 & 0 \end{pmatrix}, &
\rho^{\left(E_{g/u}\right)}(P) &= \pm \begin{pmatrix} 1 & 0 \\ 0 & 1 \end{pmatrix}.
\end{aligned} \label{eq:E-rep-rho}
\end{align}

\begin{table}[t!]
\caption{Spin-orbit matrices $\Gamma_a$ categorized according to the transformation rule~\eqref{eq:transf-Gamma} and (anti-)symmetry~\eqref{eq:Gamma-sym}.
The irrep subscript $g$ means even under parity.
The irrep superscripts $\pm$ are the values of $s$ [Eq.~\eqref{eq:Gamma-sym}], so $+$ ($-$) means that the matrices are (anti-)symmetric under transposition.
The matrices are written in terms of $[\mu, \nu] \equiv \Lambda_{\mu} \otimes \Pauli{\nu} (\iu \Pauli{y})$, where the Gell-Mann matrices $\Lambda_{\mu}$ have been listed in Appendix~\ref{sec:GM-mat}.
Subtractions of pairs $[\mu, \nu]$ represent subtractions of the respective matrices.
The two-component $\{\Gamma_1 \mid \Gamma_2\}$ transform according to the $\rho^{(E)}(g)$ given in Eq.~\eqref{eq:E-rep-rho}.
Highlighted blue are the singlet and triplet pairings with trivial orbital structures, typical of one-band SC.}
{\renewcommand{\arraystretch}{1.3}
\renewcommand{\tabcolsep}{10.6pt}
\begin{tabular}{c|c} \hline\hline
$A_{1g}^{-}$ & $\textcolor{blue}{[0,0]}$, $[2,z]$, $\textcolor{blue}{[4,0]}$, $[6,y] - [8,x]$
\\ \hline
$A_{2g}^{-}$ & $[6,x] + [8,y]$
\\ \hline
$B_{1g}^{-}$ & $[3,0]$, $[6,y] + [8,x]$
\\ \hline
$B_{2g}^{-}$ & $[1,0]$, $[6,x] - [8,y]$
\\ \hline
$E_g^{-}$ & $\left\{[2,y] \mid - [2,x]\right\}$, $\left\{[7,0] \mid - [5,0]\right\}$, $\left\{[6,z] \mid [8,z]\right\}$
\\ \hline\hline
$A_{1g}^{+}$ & $[5,y] - [7,x]$
\\ \hline
$A_{2g}^{+}$ & $\textcolor{blue}{[0,z]}$, $[2,0]$, $\textcolor{blue}{[4,z]}$, $[5,x] + [7,y]$
\\ \hline
$B_{1g}^{+}$ & $[1,z]$, $[5,y] + [7,x]$
\\ \hline
$B_{2g}^{+}$ & $[3,z]$, $[5,x] - [7,y]$
\\ \hline
$E_g^{+}$ & $\begin{matrix}
\textcolor{blue}{\left\{[0,x] \mid [0,y]\right\}}, \left\{[1,y] \mid [1,x]\right\}, \left\{[3,x] \mid - [3,y]\right\}, \\
\textcolor{blue}{\left\{[4,x] \mid [4,y]\right\}}, \left\{[5,z] \mid [7,z]\right\}, \left\{[8,0] \mid - [6,0]\right\}
\end{matrix}$ \\ \hline\hline
\end{tabular}}
\label{tab:Gamma-class}
\end{table}

To construct a $\Delta_a(\vb{k})$ that properly transforms according to Eq.~\eqref{eq:transf-Delta}, we need to combine the momentum dependence and spin-orbit structure in just the right way.
This is accomplished~\cite{Ramires2019, Kaba2019, *Kaba2019-E, Huang2019} by first separately classifying pairing wavefunctions and spin-orbit matrices (Tables~\ref{tab:d-func-class} and~\ref{tab:Gamma-class}), and then combining them according to a set of rules (Table~\ref{tab:irrep-mult}).
Let us emphasize that the emergent SC order parameter that enters Ginzburg-Landau theory belongs to the irrep determined by the total SC gap $\Delta_a(\vb{k})$ according to Eq.~\eqref{eq:transf-Delta}, and not to the irreps of its momentum or spin-orbit parts.

Pairing wavefunctions $d_a(\vb{k})$ are classified according to:
\begin{align}
d_a\!\left(R(g) \vb{k}\right) &= \sum_{b = 1}^{\dim \lambda} \rho_{ab}^{(\lambda)}(g) d_b(\vb{k}). \label{eq:transf-d}
\end{align}
All $d_a(\vb{k})$ should be made periodic, just like $\Delta_a(\vb{k})$.
If we call $\kappa_1 = a k_1$, $\kappa_2 = a k_2$, and $\kappa_3 = c k_3$, the primitive translations of a body-centered tetragonal lattice map $(\kappa_1, \kappa_2, \kappa_3)$ to $(\kappa_1 + 2 \pi, \kappa_2, \kappa_3 - 2 \pi)$, $(\kappa_1, \kappa_2 + 2 \pi, \kappa_3 - 2 \pi)$, and $(\kappa_1, \kappa_2, \kappa_3 + 4 \pi)$.
Conventionally, we also make $d_a(\vb{k})$ real so that TRSB is seen through imaginary coefficients preceding $d_a(\vb{k})$.
Examples of pairing wavefunctions are provided in Table~\ref{tab:d-func-class}.

\begin{table*}[t!]
\caption{Direct sum decompositions of the direct products between irreps of $D_{4h}$.
Since all $\Gamma$ irreps are even, the parity of the $d$ irrep and direct product irrep are the same so we have suppressed their $g/u$ subscripts.
All $E$ irreps transform according to the same 2D representation~\eqref{eq:E-rep-rho}.}
{\renewcommand{\arraystretch}{1.3}
\renewcommand{\tabcolsep}{10.0pt}
\begin{tabular}{|c"c|c|c|c|c|} \hline
$\otimes$ & $A_1\{d\}$ & $A_2\{d\}$ & $B_1\{d\}$ & $B_2\{d\}$ & $E\{d_1 \mid d_2\}$
\\ \thickhline
$A_{1g}\{\Gamma\}$ & $A_1\{\Gamma d\}$ & $A_2\{\Gamma d\}$ & $B_1\{\Gamma d\}$ & $B_2\{\Gamma d\}$ & $E\{\Gamma d_1 \mid \Gamma d_2\}$
\\ \hline
$A_{2g}\{\Gamma\}$ & $A_2\{\Gamma d\}$ & $A_1\{\Gamma d\}$ & $B_2\{\Gamma d\}$ & $B_1\{\Gamma d\}$ & $E\{\Gamma d_2 \mid - \Gamma d_1\}$
\\ \hline
$B_{1g}\{\Gamma\}$ & $B_1\{\Gamma d\}$ & $B_2\{\Gamma d\}$ & $A_1\{\Gamma d\}$ & $A_2\{\Gamma d\}$ & $E\{\Gamma d_1 \mid - \Gamma d_2\}$
\\ \hline
$B_{2g}\{\Gamma\}$ & $B_2\{\Gamma d\}$ & $B_1\{\Gamma d\}$ & $A_2\{\Gamma d\}$ & $A_1\{\Gamma d\}$ & $E\{\Gamma d_2 \mid \Gamma d_1\}$
\\ \hline
$E_g\begin{Bmatrix}\Gamma_1 \\ \Gamma_2\end{Bmatrix}$
& $E\begin{Bmatrix}\Gamma_1 d \\ \Gamma_2 d\end{Bmatrix}$
& $E\begin{Bmatrix}\Gamma_2 d \\ - \Gamma_1 d\end{Bmatrix}$
& $E\begin{Bmatrix}\Gamma_1 d \\ - \Gamma_2 d\end{Bmatrix}$
& $E\begin{Bmatrix}\Gamma_2 d \\ \Gamma_1 d\end{Bmatrix}$
& $\begin{matrix}
A_1\{\Gamma_1 d_1 + \Gamma_2 d_2\} \oplus A_2\{\Gamma_1 d_2 - \Gamma_2 d_1\} \\
\oplus B_1\{\Gamma_1 d_1 - \Gamma_2 d_2\} \oplus B_2\{\Gamma_1 d_2 + \Gamma_2 d_1\}
\end{matrix}$ \\ \hline
\end{tabular}}
\label{tab:irrep-mult}
\end{table*}

When it comes to spin-orbit matrices $\Gamma_a$, notice that $U(P) = \one$ leaves the matrix part of Eq.~\eqref{eq:transf-Delta} invariant.
This means that all spin-orbit matrices are even.\textsuperscript{\citenum{Note11}}
We classify them according to:
\begin{align}
U^{\dag}(g) \Gamma_a U^{*}(g) &= \sum_{b = 1}^{\dim \lambda} \rho_{ab}^{(\lambda)}(g) \Gamma_b, \label{eq:transf-Gamma}
\end{align}
where $U(g) = M(g) \otimes S(g)$.
Given the transposition in Eq.~\eqref{eq:exch-prop}, it is natural to further categorize $\Gamma_a$ according to (anti-)symmetry:
\begin{align}
\Gamma_a^{\intercal} &= s \Gamma_a, \label{eq:Gamma-sym}
\end{align}
where $s \in \{\pm 1\}$.
We shall also ensure time-reversal invariance:
\begin{align}
(\one \otimes \iu \Pauli{y}) \Gamma_a^{*} (\one \otimes \iu \Pauli{y}) &= \Gamma_a^{\intercal}, \label{eq:Gamma-TRI}
\end{align}
so that TRSB manifests itself through imaginary prefactors.
As the basis of the orbital part of $\Gamma_a$, we use Gell-Mann matrices $\Lambda_{\mu}$ [Appendix~\ref{sec:GM-mat}].
The spin-orbit matrices we write in terms of these:
\begin{align}
\Gamma_a &\sim \sum \Lambda_{\mu} \otimes \Pauli{\nu} (\iu \Pauli{y}).
\end{align}
Given that $\Lambda_{\mu}^{\dag} = \Lambda_{\mu}$, written thusly $\Gamma_a$ automatically satisfy time-reversal invariance~\eqref{eq:Gamma-TRI}.
In three-band systems, there are in total $4 \cdot 3^2 = 36$ possible $\Gamma_a$, of which $15$ are antisymmetric and $21$ are symmetric.
The categorization of all $\Lambda_{\mu} \otimes \Pauli{\nu} (\iu \Pauli{y}) \equiv [\mu, \nu]$ is given in Table~\ref{tab:Gamma-class}.

\footnotetext[11]{ 
Odd spin-orbit matrices occur when the conduction bands mix parities, as happens (e.g.) for topological band structures.
}

SC gap matrices $\Delta(\vb{k})$ are constructed by combining pairing wavefunctions $d_a(\vb{k})$ and spin-orbit matrices $\Gamma_a$.
Because of the exchange property~\eqref{eq:exch-prop}, we may only combine even $d_a(\vb{k})$ with antisymmetric $\Gamma_a$, or odd $d_a(\vb{k})$ with symmetric $\Gamma_a$.
Now consider a $\{d_a(\vb{k})\} \in \lambda_{d}$ and $\{\Gamma_a\} \in \lambda_{\Gamma}$, where $\lambda_{d}$ and $\lambda_{\Gamma}$ are irreps.
The object $\Delta_{ab}(\vb{k}) \equiv \Gamma_a d_b(\vb{k})$ then transforms according to the $\lambda_{\Gamma} \otimes \lambda_{d}$ representation:
\begin{equation}
\begin{aligned}
&U^{\dag}(g) \Delta_{ab}\!\left(R(g) \vb{k}\right)  U^{*}(g) = \\
&\hspace{1cm} = \sum_{a' = 1}^{\dim \lambda_{\Gamma}} \sum_{b' = 1}^{\dim \lambda_{d}} \rho_{aa'}^{(\lambda_{\Gamma})}(g) \rho_{bb'}^{(\lambda_{d})}(g) \Delta_{a'b'}(\vb{k}).
\end{aligned}
\end{equation}
Since we want to construct SC gap matrices that transform according to \emph{irreducible} representations [Eq.~\eqref{eq:transf-Delta}], we decomposed $\Delta_{ab}(\vb{k})$ into irreducible parts with the help of Table~\ref{tab:irrep-mult}.
The most general $\{\Delta_a(\vb{k})\}$ belonging to irrep $\lambda_{\Delta}$ is then given by a sum over all possible $\{d_a(\vb{k})\} \in \lambda_{d}$ and $\{\Gamma_a\} \in \lambda_{\Gamma}$ such that $\lambda_{\Delta} \in \lambda_{\Gamma} \otimes \lambda_{d}$.

For example, let us construct SC gap matrices belonging to $B_{1g}$.
In Table~\ref{tab:irrep-mult} every row has a $B_1$, meaning antisymmetric $\Gamma_a$ belonging to every irrep could be used.
Combining $[0, 0] = \Lambda_0 (\iu \Pauli{y}) \in A_{1g}^{-}$ and $\cos \kappa_1 - \cos \kappa_2 \in B_{1g}$ gives a $\Delta(\vb{k}) = \Lambda_0 (\iu \Pauli{y}) \left(\cos \kappa_1 - \cos \kappa_2\right) \in B_{1g}$, but so do many others:
\begin{align*}
A_{1g}^{-} \otimes B_{1g}\colon &\quad \left(\Lambda_6 \Pauli{y} - \Lambda_8 \Pauli{x}\right) (\iu \Pauli{y}) \left(\cos \kappa_1 - \cos \kappa_2\right), \\
A_{2g}^{-} \otimes B_{2g}\colon &\quad \left(\Lambda_6 \Pauli{x} + \Lambda_8 \Pauli{y}\right) (\iu \Pauli{y}) \sin \kappa_1 \sin \kappa_2, \\
B_{1g}^{-} \otimes A_{1g}\colon &\quad \Lambda_3 (\iu \Pauli{y}) \cos \kappa_1 \cos \kappa_2, \\
B_{2g}^{-} \otimes A_{2g}\colon &\quad \Lambda_1 (\iu \Pauli{y}) \left(\cos \kappa_1 - \cos \kappa_2\right) \sin \kappa_1 \sin \kappa_2, \\
E_g^{-} \otimes E_g\colon &\quad \Lambda_2 \left(\Pauli{x} \sin \kappa_1 - \Pauli{y} \sin \kappa_2\right) (\iu \Pauli{y}) \sin \kappa_3,
\end{align*}
etc. The most general $\Delta(\vb{k}) \in B_{1g}$ is a linear superposition of all of these.

\section{Van Hove line-gapping SC states} \label{sec:App-pics}
\begin{figure}[t]
\includegraphics[width=0.9\columnwidth]{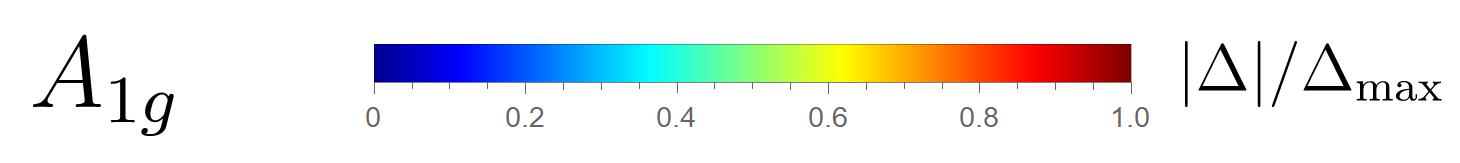}
\subfloat[$\Delta_{\vb{k}} = \Lambda_0 (\iu \Pauli{y})$]{\includegraphics[width=\columnwidth]{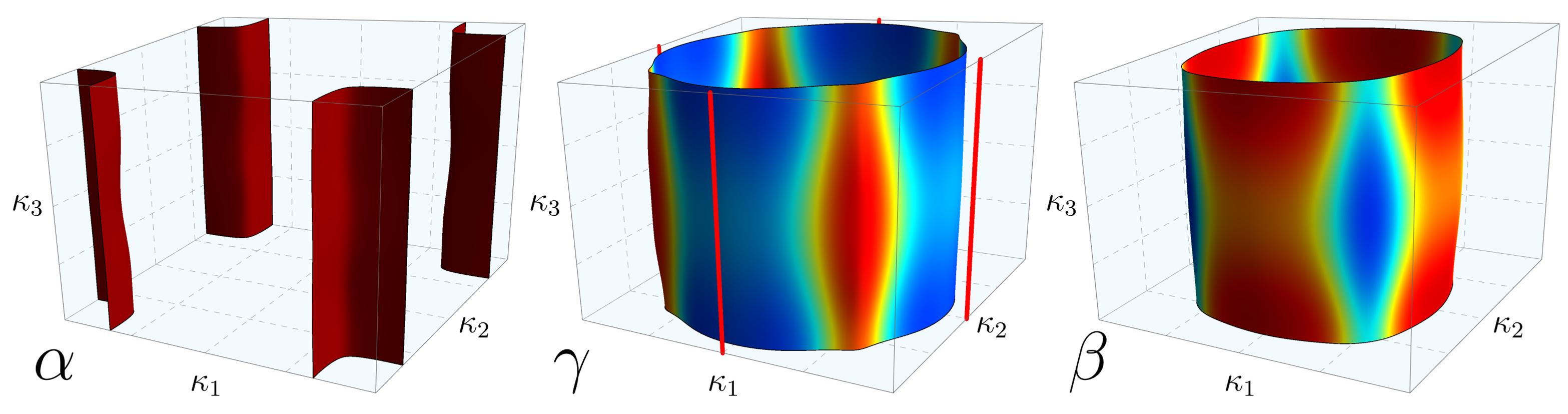}} \\
\subfloat[$\Delta_{\vb{k}} = \Lambda_2 \Pauli{3} (\iu \Pauli{y})$]{\includegraphics[width=\columnwidth]{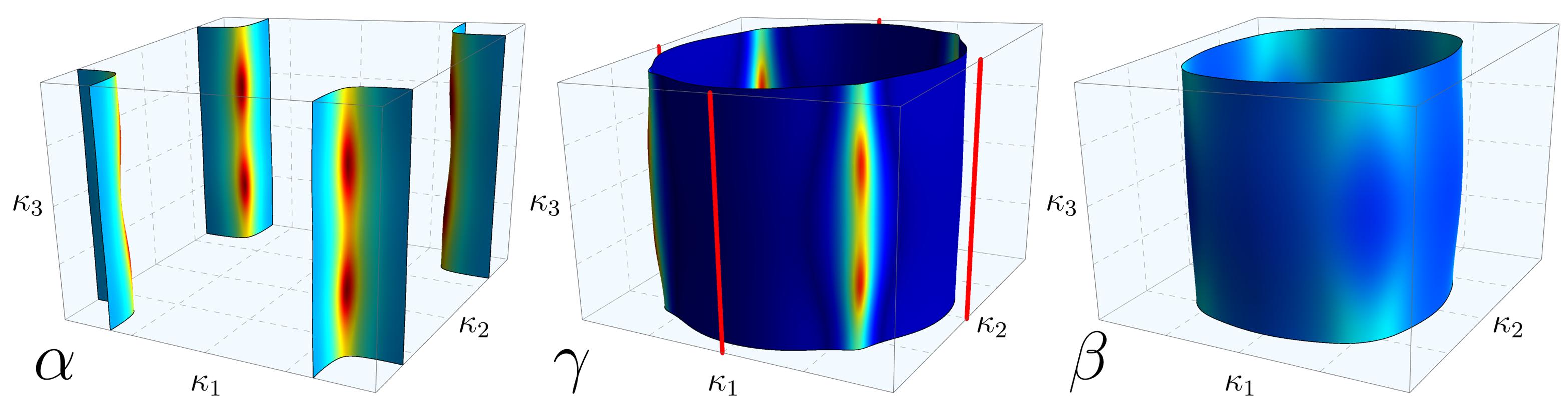}} \\
\subfloat[$\Delta_{\vb{k}} = \Lambda_4 (\iu \Pauli{y})$]{\includegraphics[width=\columnwidth]{figures/SCamplitudes/SCamplitudes_A1g-_G3_x_A1g_d1}} \\
\subfloat[$\Delta_{\vb{k}} = \left(\Lambda_6 \Pauli{y} - \Lambda_8 \Pauli{x}\right) (\iu \Pauli{y})$]{\includegraphics[width=\columnwidth]{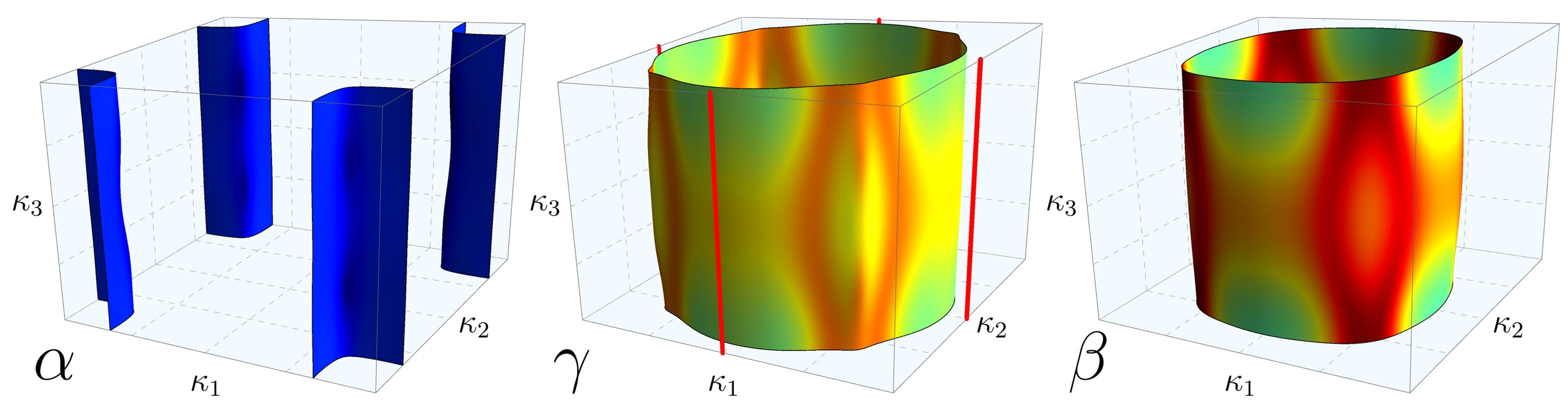}} \\
\subfloat[$\Delta_{\vb{k}} = \Lambda_3 (\iu \Pauli{y}) \left(\cos \kappa_1 - \cos \kappa_2\right)$]{\includegraphics[width=\columnwidth]{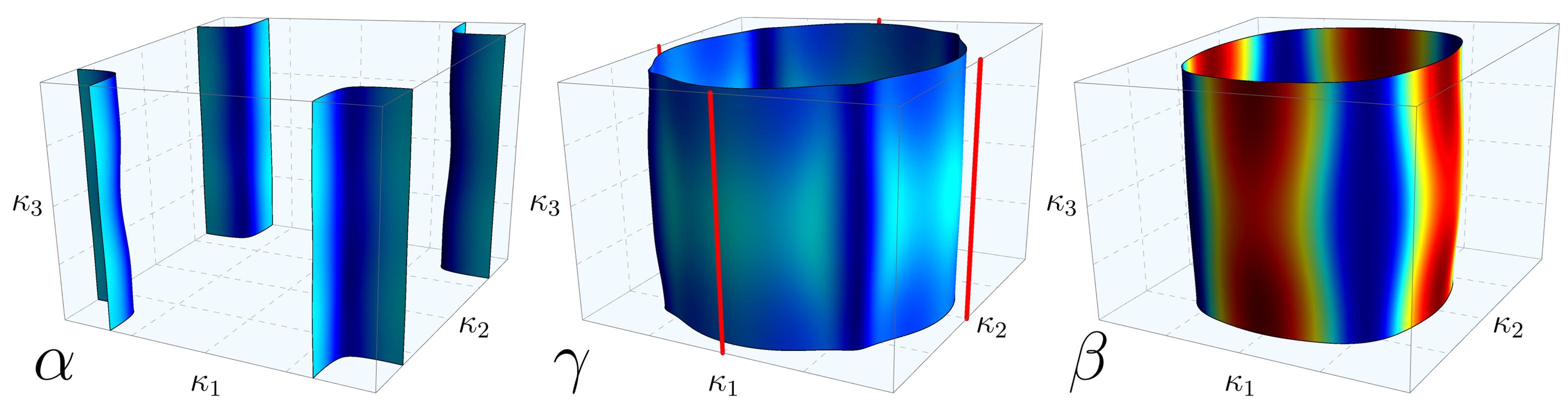}} \\
\subfloat[$\Delta_{\vb{k}} = \left(\Lambda_6 \Pauli{y} + \Lambda_8 \Pauli{x}\right) (\iu \Pauli{y}) \left(\cos \kappa_1 - \cos \kappa_2\right)$]{\includegraphics[width=\columnwidth]{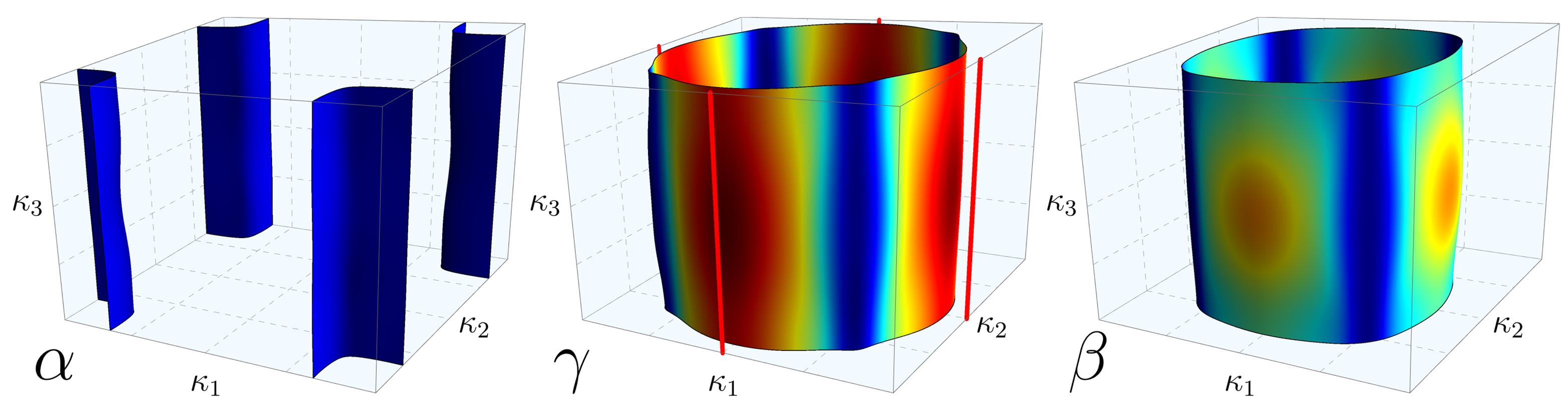}}
\caption{Projections onto the Fermi sheets of a number of Van Hove line-gapping SC states $\Delta_{\vb{k}}$ belonging to the $A_{1g}$ irrep. See the text for details.}
\label{fig:A1g-gaps}
\end{figure}

\begin{figure}[t]
\includegraphics[width=0.9\columnwidth]{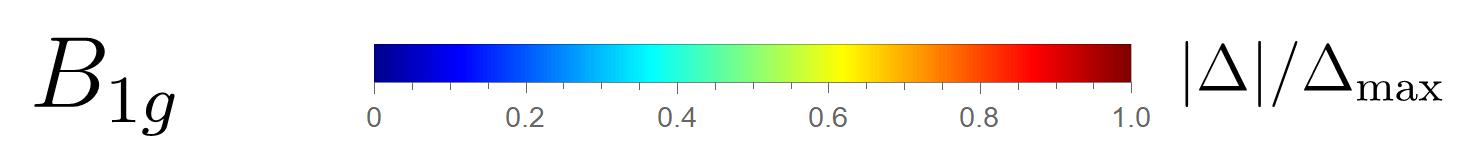}
\subfloat[$\Delta_{\vb{k}} = \Lambda_0 (\iu \Pauli{y}) \left(\cos \kappa_1 - \cos \kappa_2\right)$]{\includegraphics[width=\columnwidth]{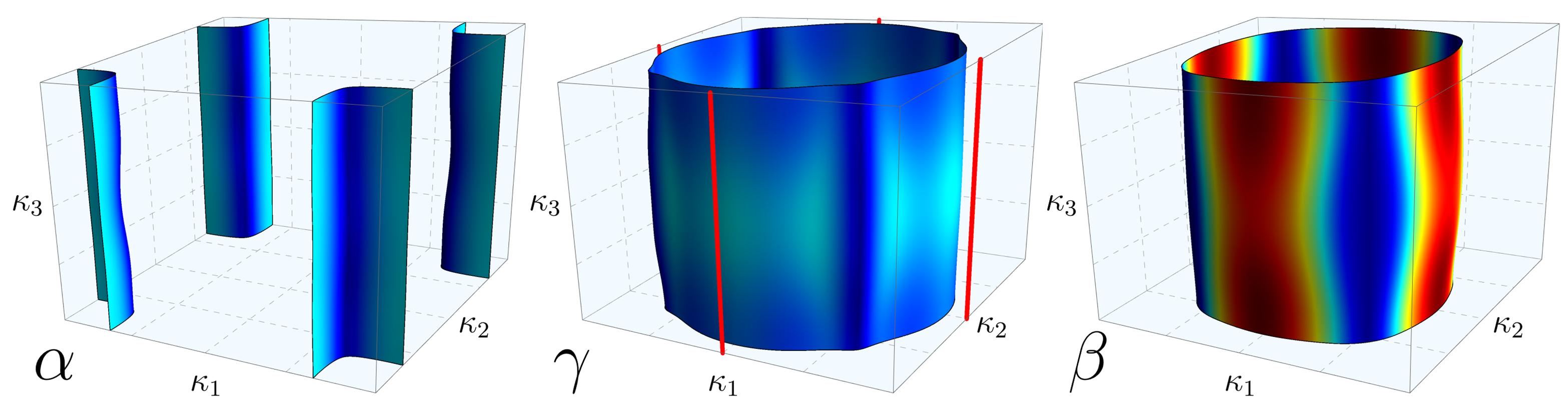}} \\
\subfloat[$\Delta_{\vb{k}} = \Lambda_2 \Pauli{3} (\iu \Pauli{y}) \left(\cos \kappa_1 - \cos \kappa_2\right)$]{\includegraphics[width=\columnwidth]{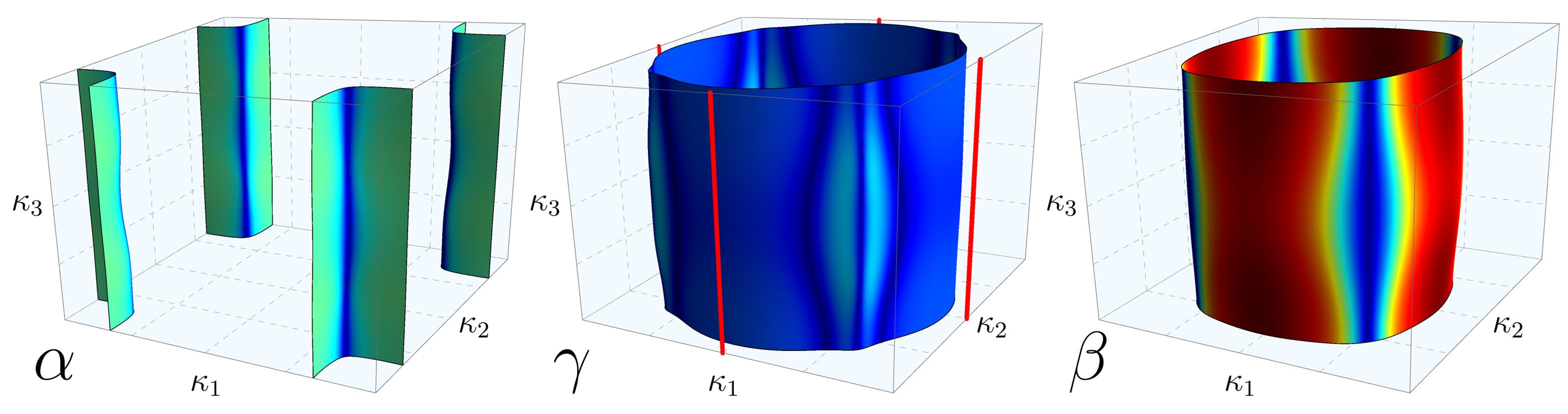}} \\
\subfloat[$\Delta_{\vb{k}} = \Lambda_4 (\iu \Pauli{y}) \left(\cos \kappa_1 - \cos \kappa_2\right)$]{\includegraphics[width=\columnwidth]{figures/SCamplitudes/SCamplitudes_A1g-_G3_x_B1g_d2}} \\
\subfloat[$\Delta_{\vb{k}} = \left(\Lambda_6 \Pauli{y} - \Lambda_8 \Pauli{x}\right) (\iu \Pauli{y}) \left(\cos \kappa_1 - \cos \kappa_2\right)$]{\includegraphics[width=\columnwidth]{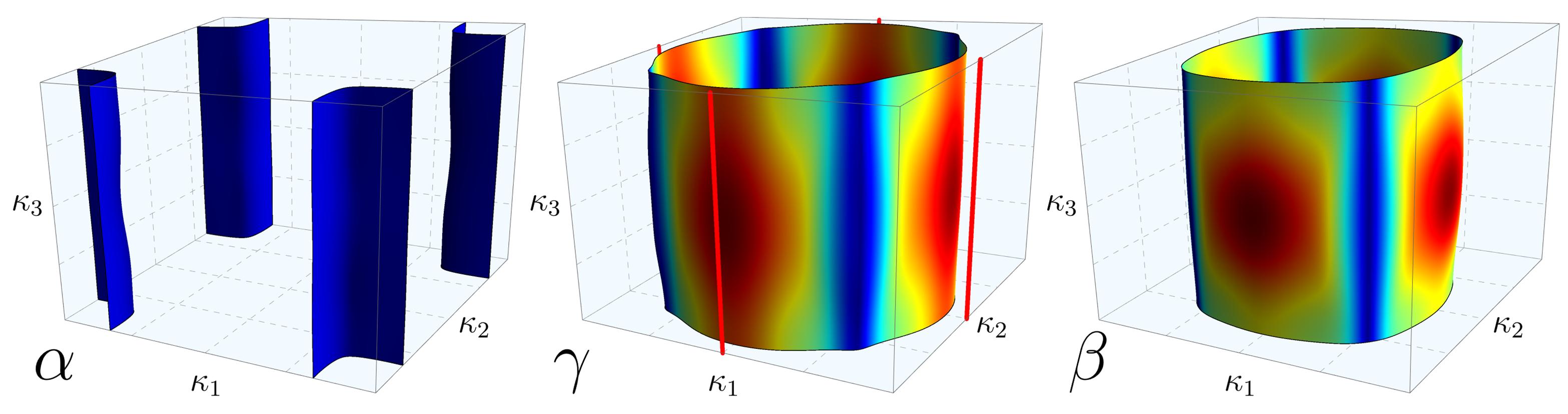}} \\
\subfloat[$\Delta_{\vb{k}} = \Lambda_3 (\iu \Pauli{y})$]{\includegraphics[width=\columnwidth]{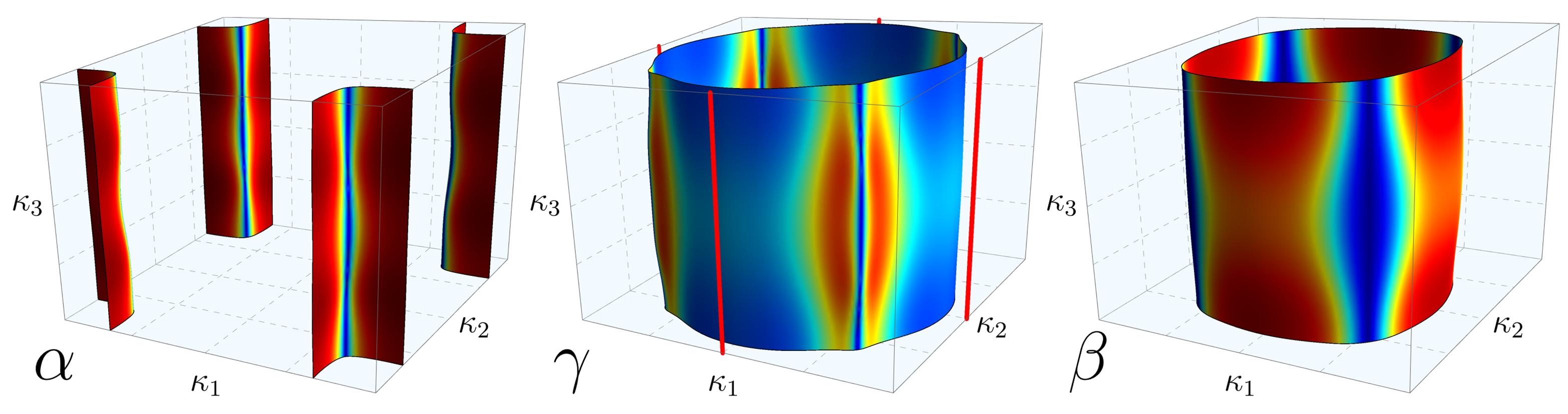}} \\
\subfloat[$\Delta_{\vb{k}} = \left(\Lambda_6 \Pauli{y} + \Lambda_8 \Pauli{x}\right) (\iu \Pauli{y})$]{\includegraphics[width=\columnwidth]{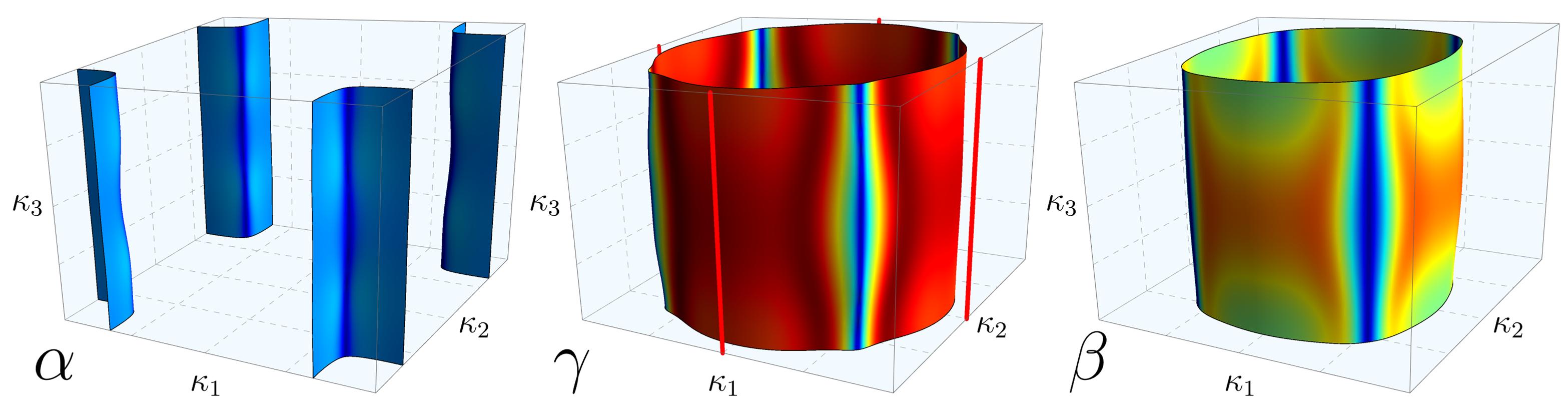}}
\caption{Projections onto the Fermi sheets of a number of Van Hove line-gapping SC states $\Delta_{\vb{k}}$ belonging to the $B_{1g}$ irrep. See the text for details.}
\label{fig:B1g-gaps}
\end{figure}

\begin{figure}[t]
\includegraphics[width=0.9\columnwidth]{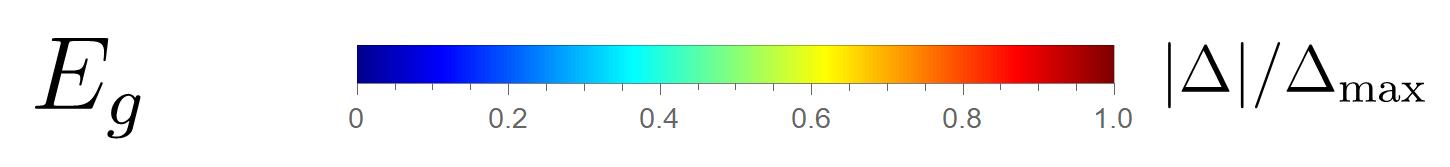}
\subfloat[$\Delta_{\vb{k}} = \Lambda_0 (\iu \Pauli{y}) d_{(x \pm \iu y) z}(\vb{k})$]{\includegraphics[width=\columnwidth]{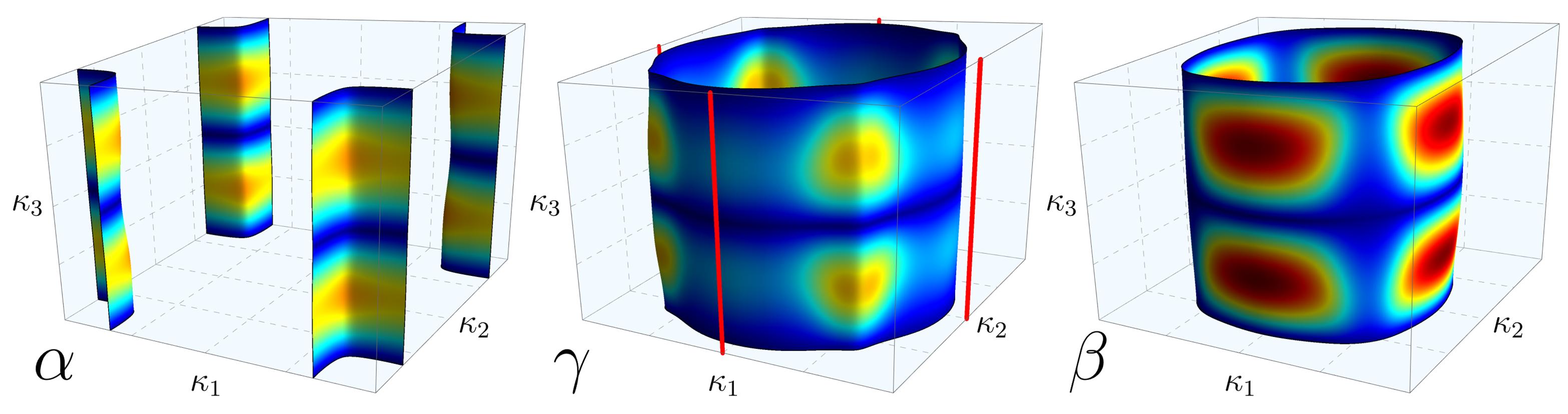}} \\
\subfloat[$\Delta_{\vb{k}} = \Lambda_2 \Pauli{3} (\iu \Pauli{y}) d_{(x \pm \iu y) z}(\vb{k})$]{\includegraphics[width=\columnwidth]{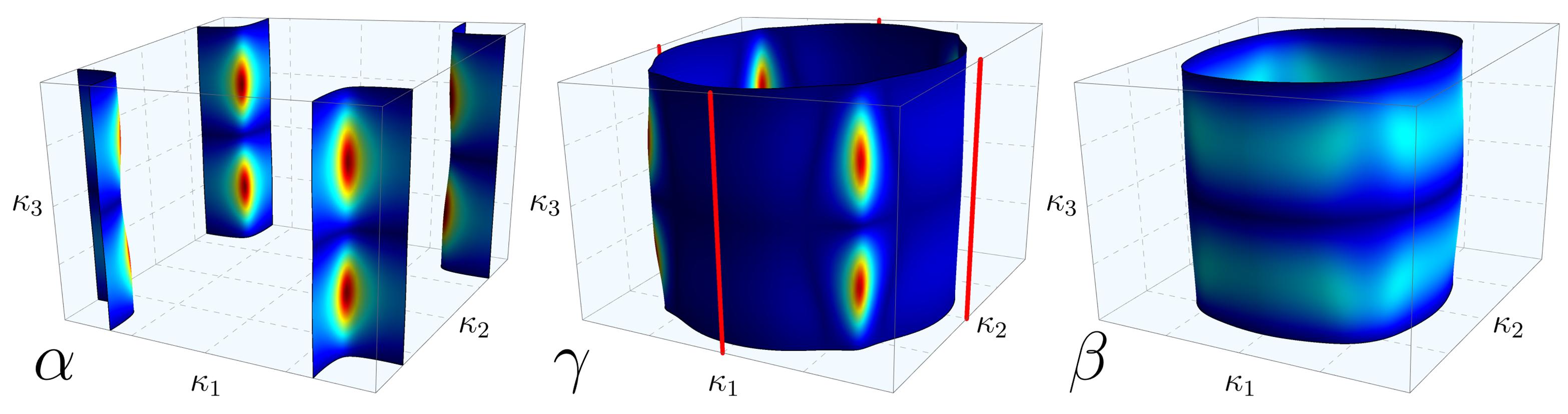}} \\
\subfloat[$\Delta_{\vb{k}} = \Lambda_4 (\iu \Pauli{y}) d_{(x \pm \iu y) z}(\vb{k})$]{\includegraphics[width=\columnwidth]{figures/SCamplitudes/SCamplitudes_A1g-_G3_x_Eg_d3}} \\
\subfloat[$\Delta_{\vb{k}} = \left(\Lambda_6 \Pauli{y} - \Lambda_8 \Pauli{x}\right) (\iu \Pauli{y}) d_{(x \pm \iu y) z}(\vb{k})$]{\includegraphics[width=\columnwidth]{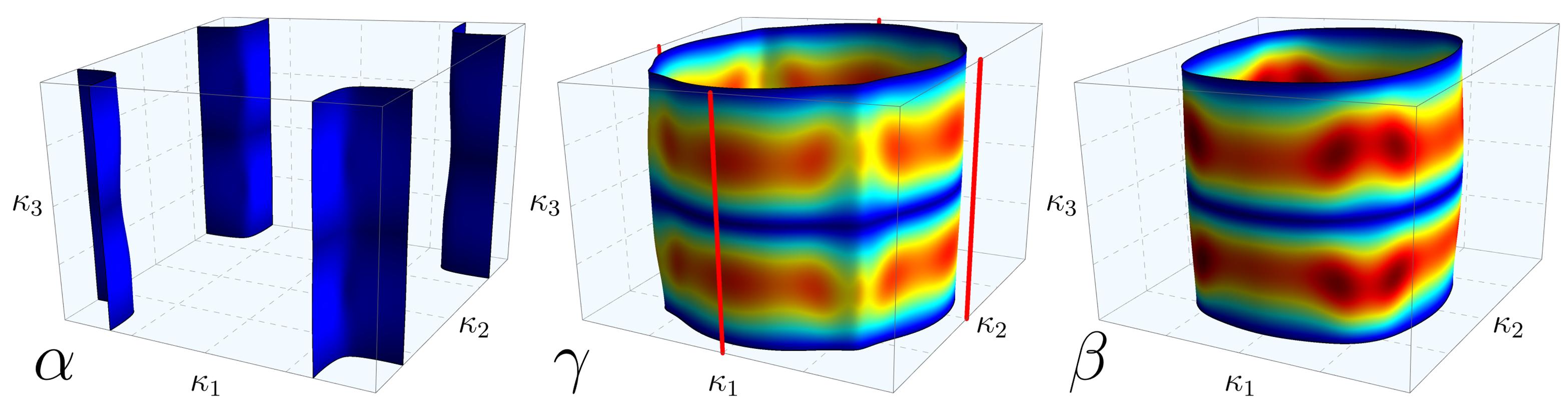}} \\
\subfloat[$\Delta_{\vb{k}} = \Lambda_3 (\iu \Pauli{y}) d_{(x \pm \iu y) z}(\vb{k})$]{\includegraphics[width=\columnwidth]{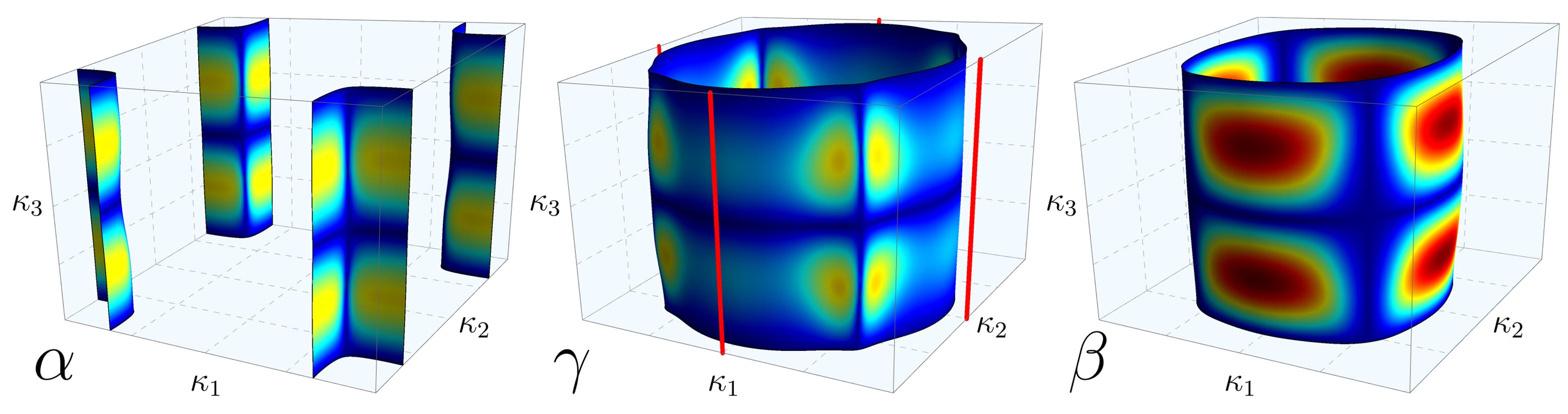}} \\
\subfloat[$\Delta_{\vb{k}} = \left(\Lambda_6 \Pauli{y} + \Lambda_8 \Pauli{x}\right) (\iu \Pauli{y}) d_{(x \pm \iu y) z}(\vb{k})$]{\includegraphics[width=\columnwidth]{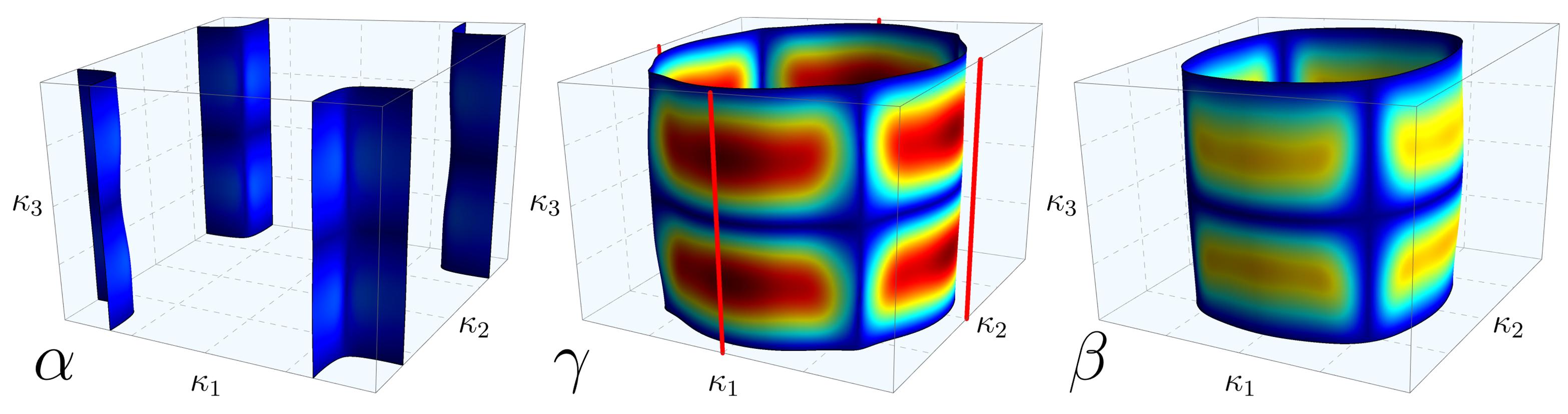}}
\caption{Projections onto the Fermi sheets of a number of chiral Van Hove line-gapping SC states $\Delta_{\vb{k}}$ belonging to the $E_{g}$ irrep. See the text for details.}
\label{fig:Eg-gaps}
\end{figure}

In Figures~\ref{fig:A1g-gaps}, \ref{fig:B1g-gaps}, and~\ref{fig:Eg-gaps}, we have plotted the Fermi surface-projections of a number of Van Hove line-gapping even SC states from Table~\ref{tab:main-result}.
These have been constructed by combining the six $A_{1g}^{-}$ and $B_{1g}^{-}$ spin-orbit matrices [Table~\ref{tab:Gamma-class}] with the lowest order $A_{1g}$, $B_{1g}$, and $E_{g}$ pairing wavefunctions [Table~\ref{tab:d-func-class}].
$\Delta_{\vb{k}}$ constructed from the highly suppressed $E_g^{-}$ spin-orbit matrices aren't shown.
Of all the possible superpositions in the case of $E_g$ pairing, we have shown the chiral ones as they are the most interesting because of the various evidence~\cite{Luke1998, Luke2000, Higemoto2014, Grinenko2021-unaxial, Grinenko2021-isotropic, Xia2006, Kapitulnik2009} indicating TRSB.
The most general Van Hove line-gapping $\Delta_{\vb{k}}$ belonging to $A_{1g}$, $B_{1g}$, or chiral $E_g$ is a superposition of the shown ones, plus higher order harmonics.
$\kappa_1 = a k_1 \in [- \pi, \pi]$, $\kappa_2 = a k_2 \in [- \pi, \pi]$, $\kappa_3 = c k_3 \in [- 2 \pi, 2 \pi]$, and $d_{(x \pm \iu y) z}(\vb{k}) = \left(\sin\frac{\kappa_1}{2} \cos\frac{\kappa_2}{2} \pm \iu \cos\frac{\kappa_1}{2} \sin\frac{\kappa_2}{2}\right) \sin\frac{\kappa_3}{2}$.
In the $\gamma$ sheet plots, the Van Hove lines $\left(\pm \frac{\pi}{a}, 0, k_3\right)$ and $\left(0, \pm \frac{\pi}{a}, k_3\right)$ have been highlighted red.
Even though the projections of some $\Delta_{\vb{k}}$ onto the $\gamma$ band might be small (shaded blue) near the Van Hove lines (e.g., Figure~\ref{fig:A1g-gaps}~(b)), they are only exactly zero at a certain $\kappa_3$ for the $\Delta_{\vb{k}} \in E_g$ that have horizontal nodes at $\kappa_3 = 0, \pm 2 \pi$.

\clearpage

\bibliography{Sr2RuO4-references}

\end{document}